\documentclass[11pt]{article}
\usepackage{graphicx,amsmath,amssymb}
\usepackage[usenames]{color}
\usepackage[colorlinks=true, urlcolor=navyblue, linkcolor=navyblue, citecolor=navyblue]{hyperref}
\usepackage{epstopdf}
\usepackage{enumerate}
\usepackage{enumitem}  
\setlist{nosep}
\usepackage{appendix}
\usepackage{lineno}
\usepackage{setspace}
\usepackage{wrapfig}
\usepackage{lipsum}
\usepackage{placeins}
\usepackage{bbold}
\usepackage{xfrac}
\usepackage{xcolor}
\usepackage{float}

\definecolor{navyblue}{rgb}{0,0.08,0.45}

\begin{document}

\title{Measurement of the high-energy contribution \\ 
to the Gerasimov-Drell-Hearn sum rule}

\date{}

\maketitle

\vspace{-40pt}

\centerline{
M.~M. Dalton$^{*}$, A.~Deur$^{* \ddagger}$, C.~D. Keith}
\centerline {\it Thomas Jefferson National Accelerator Facility, Newport News, VA 23606, USA}
\vspace{10pt}

\centerline{S.~\v{S}irca$^{*}$}
\centerline {\it University of Ljubljana, Ljubljana, Slovenia}
\vspace{10pt}

\centerline{J.~Stevens$^{*}$}
\centerline {\it William \& Mary, Williamsburg, Virginia 23187, USA}
\vspace{10pt}

\vspace{5pt}
\centerline{Endorsed by the GlueX Collaboration}

\vspace{5pt}
\noindent$^*$ Spokesperson \\
$\ddagger$ Contact \\

\noindent{\Large\bf Abstract}

\bigskip

\noindent We propose to measure the high-energy behavior of the integrand of the Gerasimov-Drell-Hearn (GDH) sum rule on the proton and the neutron up to 12 GeV.  
The convergence of the GDH integral will be investigated for the first time and to high precision.  
The validity of the GDH sum rule on the neutron will be accurately tested for the first time, while for the proton the uncertainty will be improved by 25\% relative.  
The data will allow precision testing of Regge phenomenology in the polarized domain. The $a_1$ and $f_1$ Regge trajectory intercepts will be obtained to an order of magnitude higher precision than the current best estimates.  
The data will also contribute to the determination of the real and imaginary parts of the spin-dependent Compton amplitude, the polarizability correction to hyperfine splitting in 
hydrogen, and to studying the transition between polarized DIS and diffractive regimes.  

The experiment will require a circularly polarized photon beam (produced from a longitudinally polarized electron beam) with a flux approximately \sfrac{1}{3} of the GlueX-II experiment E12-13-003.  
The experiment will run in two configurations which require two different CEBAF beam energies.  
A new longitudinal polarized proton and deuteron target will be needed in Hall D.  
The experiment will require 21 PAC days at the nominal CEBAF energy and another 12 PAC days at an energy \sfrac{1}{3} to \sfrac{1}{2} of the nominal.

\newpage
\label{endorsement}
\begin{figure}[H]
\centering
\hspace{-1.cm}
\includegraphics[width=0.99\textwidth]{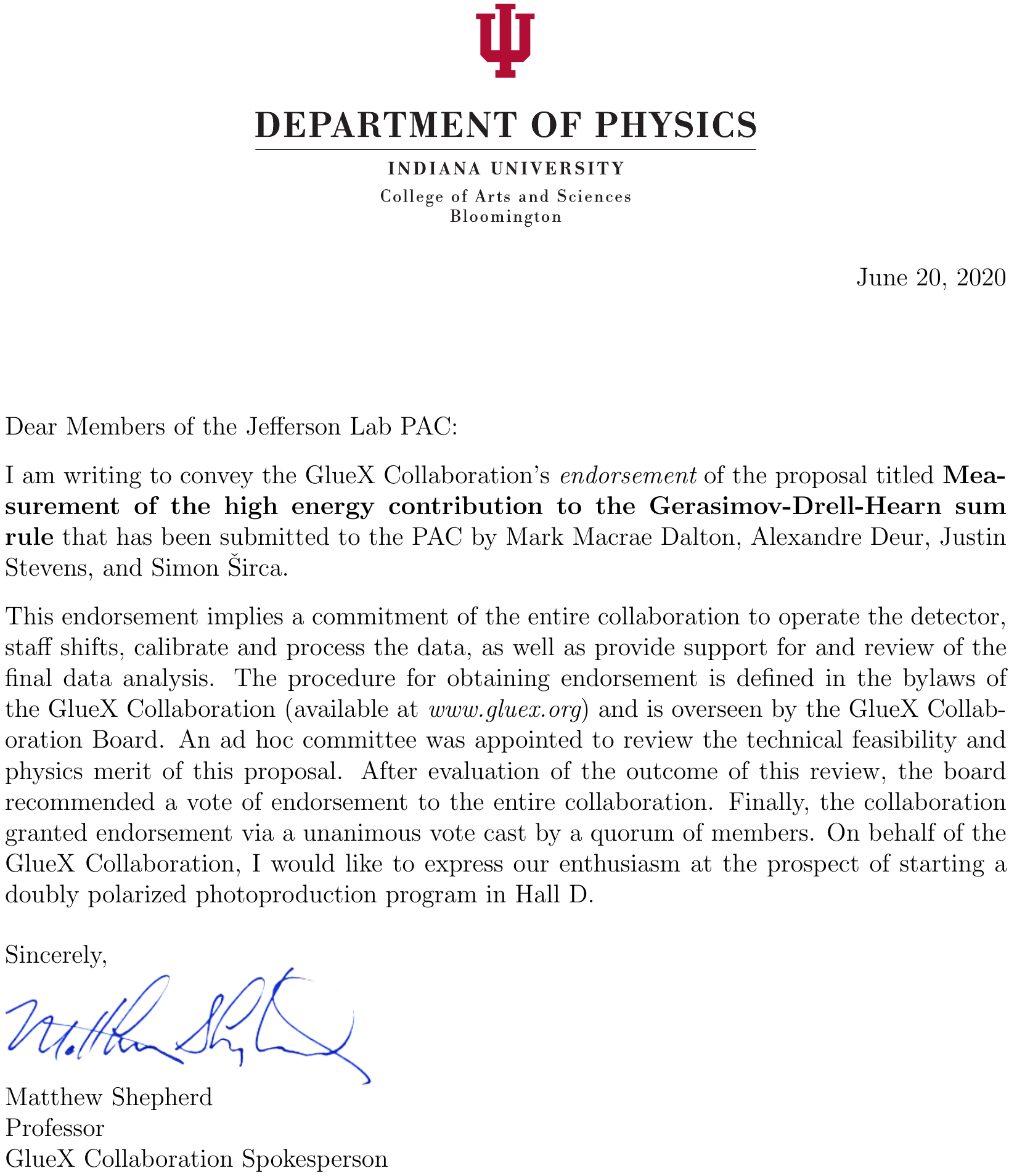}
\vspace{-0.0cm}
\end{figure}
\newpage
\tableofcontents
\newpage

\section{Introduction}

\subsection{Executive summary}

\noindent We propose to measure the high-energy behavior of the integrand 
of the Gerasimov-Drell-Hearn (GDH) sum rule, i.~e.~of the doubly 
polarized total photoproduction cross-section asymmetry.  
The high-energy domain is where a failure of the GDH sum rule may occur
for a number of reasons elaborated below, and such behavior 
would indicate unknown structural features or dynamic processes
in the nucleon.
The data will also improve significantly the precision at
which the sum rule is tested on the proton, and offer a test of comparable
accuracy for the neutron, which is not tested at present.

Independently of the sum rule study, the measurement will
investigate QCD in an energy domain where its phenomenology 
is unknown when spin degrees of freedom are explicit.  
The experiment will thus provide a baseline for the EIC's study 
of the transition between polarized deep inelastic scattering (DIS) 
and polarized diffractive regimes. In particular, 
our data will help to clarify a discrepancy between fits of the photoproduction 
and DIS world data and the corresponding theoretical expectations, which give 
conflicting predictions for the power-law dependence of the GDH integrand. 

The experiment will be sensitive enough to provide, for the first time, 
a precise measurement of the deuteron asymmetry in the diffractive regime. 
Chiral effective field theory will also be tested in a different regime 
than that covered by the low-$Q^2$ JLab spin sum rule program.  
Finally, our measurements will constrain the polarizability contribution 
to the hydrogen hyperfine splitting.

We propose to perform the measurement on the proton and neutron in Hall~D 
with CEBAF at 12~GeV.  Hall~D is uniquely  suited for such a measurement 
thanks to its photon tagger and its high-luminosity, large solid angle detector.
Overall, the experiment aims at providing an absolute measurement of the polarized cross-section difference
at a $\approx 5\%$ accuracy, typical for such experiments.   However, 
the key goal of the experiment---to determine the high-energy behavior 
of the GDH integrand---does not require absolute normalization and thus 
will have significantly reduced uncertainties of about $2\%$, since only point-to-point uncorrelated errors contribute.

\medskip

A shorter version of this document was submitted as a Letter of Intent 
to PAC47 \cite{LOI}. Based on it, the PAC acknowledged the feasibility 
of the experiment, considered Hall~D best suited for performing it, 
and encouraged this first step toward a comprehensive doubly-polarized 
program in Hall~D \cite{PAC47_LOI_Comments}: 
``{\sl The PAC recognizes the science case for this LOI 
and recommends preparation of a full proposal with focus on the extraction 
of the actual value of the GDH integral at high energies. The PAC would 
be pleased to see the development of ideas towards a full program with 
a circularly polarized photon beam and a polarized target in Hall~D.}''

The proposal is endorsed by the GlueX collaboration, which will support the preparation of the experiment, its run, analysis and publications related to the experiment, see the letter of endorsement page~\pageref{endorsement}.

\subsection{The Gerasimov-Drell-Hearn sum rule \label{GDH intro}}
The Gerasimov-Drell-Hearn (GDH) sum rule~\cite{Gerasimov:1965et} is a general and fundamental relation that links the anomalous magnetic moment $\kappa$ of a particle to its helicity-dependent photoproduction cross-sections:
\begin{equation}
I \equiv \int_{\nu_0}^{\infty}\frac{\Delta \sigma(\nu)}{\nu}\,\mathrm{d}\nu=\frac{4\pi^2 S \alpha_\text{em}\kappa^2}{M^2},
\label{eq:gdh}
\end{equation}
where $\nu$ is the probing photon energy, $S$ is the spin of the target particle, $M$ is its mass and $\nu_0 = m_\pi(1+m_\pi/2M)$ is the threshold energy
for pion photoproduction ($m_\pi$ is the neutral pion mass), and $\alpha_\text{em}$ is the electromagnetic coupling constant.
In our case (proton and neutron)  $S=1/2$ and 
$\Delta \sigma \equiv \sigma_{P} - \sigma_{A}$
is the difference in total photoproduction cross-sections 
($\gamma N \to X$) for which the  photon spin is parallel and anti-parallel to the target particle spin, respectively.  Note that with this relative sign definition the GDH integral is positive---the opposite convention is also seen in the literature. 

The sum rule is valid for any type of particle: nucleons, nuclei, electrons, even photons.
For the proton, the right-hand side of Eq.~(\ref{eq:gdh}) 
gives $I^p= 204.78\,\mu$b, while for the neutron, one obtains
$I^n=233.52$ $\mu$b. 

The experiment has two major thrusts, which we call \emph{convergence} and \emph{validity}.  The first thrust is to verify that the GDH integral converges to a finite value.  It is the high-energy behavior of the integrand which determines this.  Mathematically, since it is weighted by $1/\nu$, $\sigma_{P}-\sigma_{A}$ must decrease faster than $1/\log\nu$ in order for the integral to converge.  We will test this by precisely measuring the high-$\nu$ dependence of the spin-dependent cross-section difference.

The second thrust is to improve the determination of the GDH integral itself which will impose a more stringent test on the \emph{validity} of the sum rule, and thereby improve our sensitivity to physical processes that may cause a real or an apparent violation of the sum rule.

\begin{figure}[ht]
\centering
\includegraphics[width=0.6\textwidth]{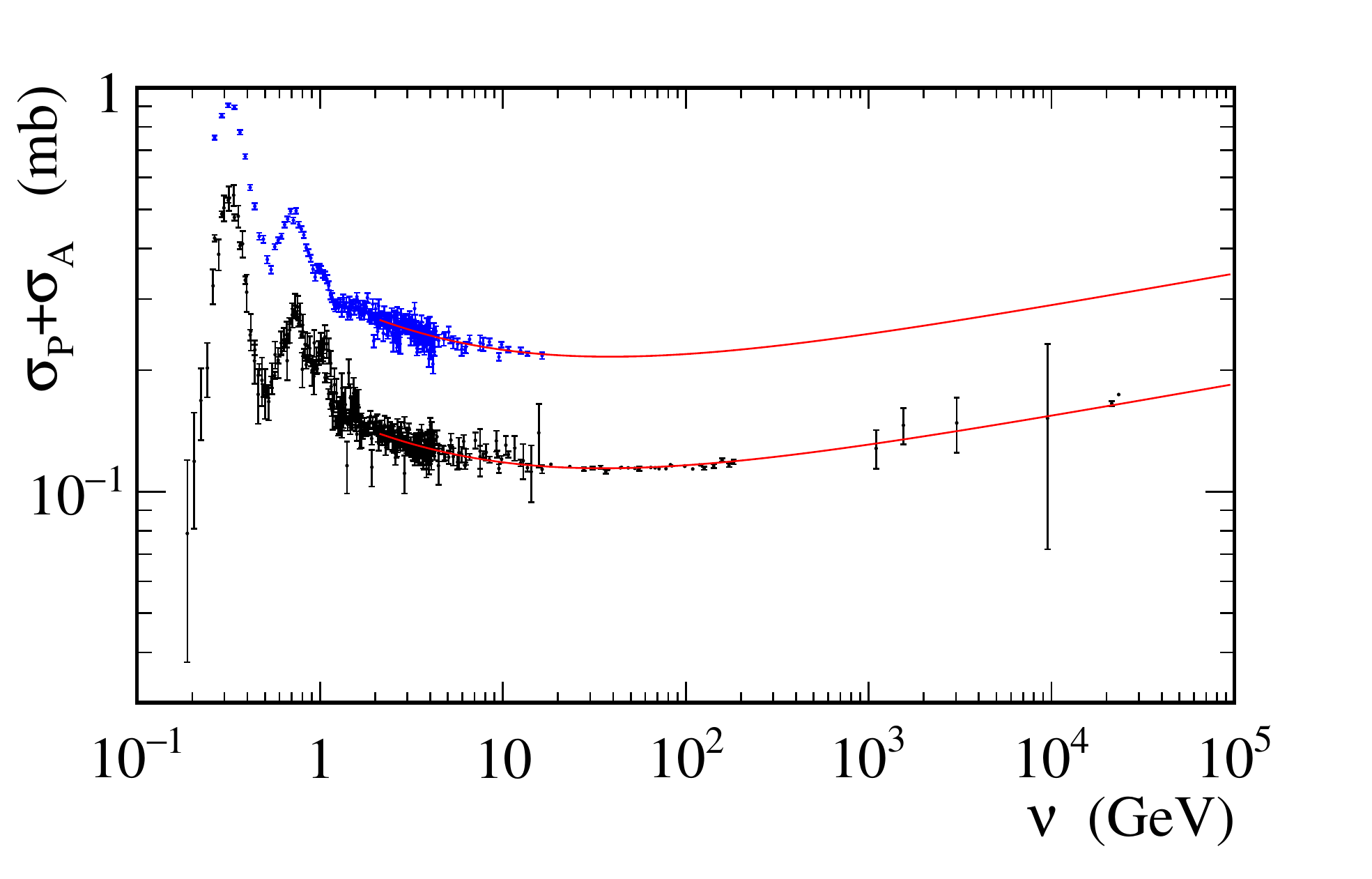}
\vspace{-0.4cm}
\caption{\label{fig:totalCS}
\small{Unpolarized total photoabsorption cross-section $ \sigma_P + \sigma_A$ for the proton (black symbols) and deuteron (blue symbols) as a function of the photon beam energy. The data are from Ref.~\cite{Tanabashi:2018oca}. The lines are Regge fits including a pomeron term proportional to $\nu^{0.08}$.
}}
\end{figure}

It is illustrative to consider the unpolarized equivalent of the GDH sum rule, 
\begin{equation}
\int_{\nu_0}^{\infty}(\sigma_{P}+\sigma_{A})\,\mathrm{d}\nu=-\frac{\alpha}{M}.
\label{eq:unpo gdh}
\end{equation}
This ``rule" itself is clearly invalid, as the spin-independent (total) cross-section is a positive definite quantity (like all cross-sections) while the sum rule sets it equal to a negative number.
In addition, the integral itself does not even converge to a finite value due to the behavior of the integrand at high energy.  Fig.~\ref{fig:totalCS} shows the spin-independent cross-section as a function of $\nu$.  The high-$\nu$ behavior can be described by  ($\sigma_{P}+\sigma_{A}) \propto \nu^{0.08}$, which does not result in a finite integral. The empirical observation of this divergence in multiple hadronic interactions, has led to the postulation of the pomeron in 1961.  This divergence and the resulting insight would not become
apparent without extending the measurements to sufficiently high energies. 

In contrast to this, the sum rule obtained by using the second order of the low-energy theorem (see Section~\ref{GDH derivation}) has an additional $1/\nu^2$ factor, yielding the Baldin sum rule~\cite{baldinSR},
\begin{equation}
\int_{\nu_0}^\infty\frac{\sigma_P + \sigma_A}{\nu^2}\,\mathrm{d}\nu=4\pi^2(\alpha_E+\beta_M),
\label{eq:baldin}
\end{equation}
where $\alpha_E$ and $\beta_M$ are the electric
and magnetic polarizabilities, respectively.  
The $1/\nu^2$ factor overcomes the convergence problem, and polarizability measurements have verified the sum rule prediction~\cite{Ahrens:2001qt,Gurevich:2017cpt}.

\subsection{High-energy behavior, Regge theory and GDH integral convergence}

A failure of the GDH integral to converge would be a startling development and would immediately indicate that some phenomenology is missing from our understanding.  We propose to measure the \emph{functional form} of the spin-dependent cross-section difference $\Delta \sigma$ at high photon energy to high precision.  This part of the measurement does not rely on any existing data and is not sensitive to many systematic uncertainties affecting measurements of absolute cross-sections.  

Regge theory predicts the cross-section at high energy to be described by the functional form given in Eq.~(\ref{eq:Regge_expect})
where the parameters must be determined from the data~\cite{Bass:1997fh}.
Our data will allow a test of Regge theory well into the region where it is expected to be applicable.
Mueller and Trueman~\cite{Mueller:1967zz} assessed that if Regge behavior holds, the spin-dependent cross-section should drop to zero faster than $1/\log^2\nu$, which would be sufficient for the GDH integral to converge.
Section~\ref{sec:ShapeSens} discusses the sensitivity of our measurement to the parameters of the presumed Regge dependence. 

In addition to the sum rule study, our measurement will investigate QCD in its diffractive scattering regime, where Regge theory is expected to describe the scattering process. It would be the first clean test for polarized photoproduction.
As signaled in the review of Ref.~\cite{Drechsel:2007sq},
this phenomenology has not been tested with spin degrees of freedom:
\emph{ ``above the resonance region, one usually invokes Regge phenomenology to argue that the integral converges [...] However, these ideas have still to be tested experimentally. [...] the real photon is essentially absorbed by coherent processes, which require interactions among the constituents such as gluon exchange between two quarks. This behavior differs from DIS, which refers to incoherent scattering off the constituents.''} 
The lack of spin-dependent tests is an important shortcoming also emphasized by Bjorken~\cite{Bjorken:1996dc}: \emph{``Polarization data has often been the graveyard of fashionable theories. If theorists had their way, they might well ban such measurements altogether out of self-protection."}
This is supported by a stark discrepancy that exists between fits of the photoproduction and DIS world data and expectations from Regge theory, see Section~\ref{impact on a_1 intercept.}.

\subsection{Validity of the sum rule}

The saturation of the integral beyond a given $\nu$ indicates 
the energy scale at which the characteristic scale
of the object structure becomes irrelevant (or its mass scale scale for a structureless object).  

For a lepton, at first order in perturbation, 
$\Delta \sigma(\nu)$ is non-zero only at $\nu$ in the vicinity of the lepton 
mass \cite{Pantforder:1998nb} (where it switches sign to ensure that the
integral yields zero).  

For a nucleon, only a single quark participates 
in a high-energy reaction and, if quarks are structureless, $\kappa_q = 0$ 
for the active quark and it does not contribute to the sum rule. 
A failure of the sum rule to saturate at its expected value by a certain $\nu$ imply 
that there remains an additional contribution at higher energy.
Measuring the sum rule to 12\,GeV allows one to bound the contributions to the nucleon structure that come from energy scales larger than $12$\,GeV.
Other possible causes that could invalidate the sum rule exist, and all involve high-$\nu$ phenomenology, see Section~\ref{sec:theory}.

Thus, while the nucleon GDH sum gets most of its contribution from the resonance regime,  the high-energy part is critical since it may reveal possible substructure or unknown structural processes.  Indeed, it is the high energy domain that would expose a failure of the sum rule.

\subsection{Target nucleons}

It is important to measure $\Delta \sigma$ on both the proton and the neutron for two reasons. 

First, we would provide two independent tests of the sum rule, since processes in the neutron may be different to those in the proton. In general, it is possible for the sum rule to appear valid for one type of target, and invalidated for other ones. For example, it has been shown that within the Standard Model, the sum rule is true for the electron, but this has no bearing on its validity for, e.g., nucleon targets.   Gathering neutron data in Hall D would be especially important because the neutron world data are not as extensive as those for the proton: they are less precise and extend only to up to $\nu$ of 1.8\,GeV, see Fig.~\ref{fig:GDH_world}. 

Second, gathering data on both nucleons allow for an isospin analysis of their high-$\nu$ behavior, as discussed next.  

\subsection{Isospin decomposition}

Regge theory suggests that at high $\nu$, 
$\Delta \sigma(\nu) \propto (\nu+M/2)^{\alpha_0-1}$~\cite{Bass:1997fh}, with $\alpha_0$ a Regge intercept.
For the isovector part, $\Delta \sigma^{p-n}\equiv\Delta \sigma^{p}-\Delta \sigma^{n}$, $\alpha_0$ should be determined by the $a_1(1260)$ meson trajectory, which is still not well known.  For the isoscalar $\Delta \sigma^{p+n}\equiv\Delta \sigma^{p}+\Delta \sigma^{n}$ part, $\alpha_0$ should be  given by the $f_1(1285)$, which is better known.  Thus, an analysis of isospin decomposition of the Regge trajectories requires an accurate measurement of the proton and the neutron.

Independent of the $f_1(1285)$ intercept value, i.e. of the question of $\nu$-dependence, the absolute normalization of $\Delta \sigma^{p+n}$ (i.e. $c_2$ in Eq.~(\ref{eq:Regge_expect})) is not precisely known. In fact, $\Delta \sigma^{p+n}$ is at present assumed to be zero in analyses since the measured asymmetry on the deuteron in the diffractive regime is consistent with zero~\cite{Bass:2000zv, Bass:2018uon}.  
This experiment will be precise enough to measure clearly, and for the first time, a non-zero polarized deuteron signal in this regime  (at least $10\sigma$ based on Regge expectations).

\subsection{Additional physics impact}

Theoretical dispersion analysis of the measured 
$\Delta \sigma(\nu)$ will yield the complex spin-dependent Compton amplitude $f_2(\nu)$ (see next section and Section~\ref{impact with f2}), and thereby tests chiral effective field theory ($\chi$EFT), the leading non-perturbative approach to QCD at low energy-momentum.  This will complement the JLab low-$Q^2$ spin sum rule experimental program that tested $\chi$EFT and showed that the description of spin observables remains a challenge for  $\chi$EFT~\cite{Deur:2018roz}. 

Finally, measuring $\Delta \sigma(\nu)$ will  provide a baseline for some of the Electron Ion Collider (EIC)  studies, as well as constrain the polarizability contribution to the hydrogen hyperfine splitting, see Sections~\ref{impact HFS} and~\ref{impact EIC}.

\subsection{List of GDH experiments \label{GDH Experiments}}

\begin{figure}[ht]
\centering
\includegraphics[width=0.75\textwidth]{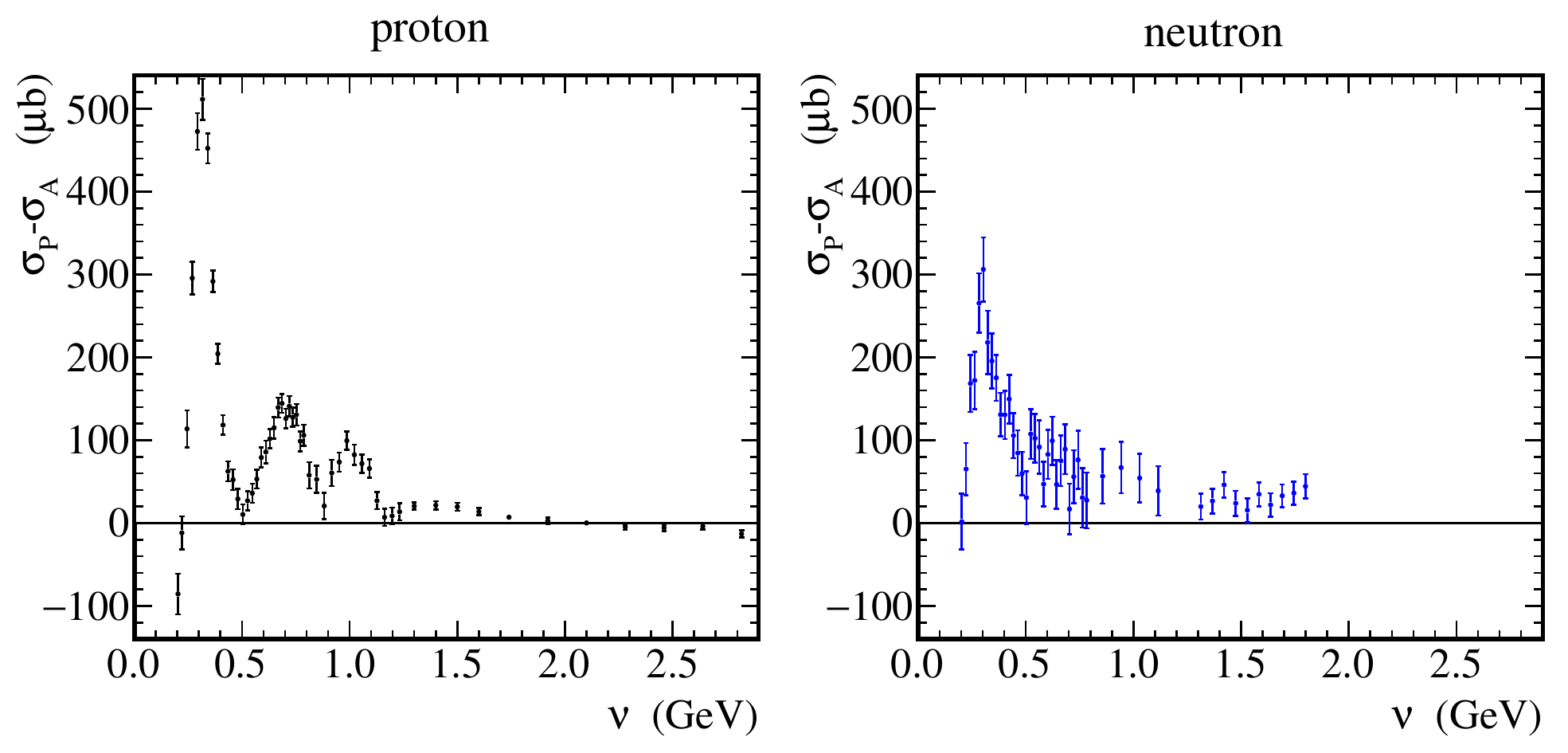}
\vspace{-0.3cm}
\caption{\label{fig:GDH_world}\small{World data of the spin-dependent cross-section difference $\Delta \sigma$ on the proton (left) and neutron (right).  Data from various experiments are combined and rebinned.}}
\end{figure}

The GDH integrand was measured at MAMI and ELSA for energies in the range
$0.2 \,\mathrm{GeV} \leq \nu \leq 2.9\,\mathrm{GeV}$ for the proton and in the $0.2 \,\mathrm{GeV} \leq \nu \leq 1.8\,\mathrm{GeV}$ range for the neutron, see Fig.~\ref{fig:GDH_world}. 
Partial contributions to the sum rule (from individual channels) were also measured 
at LEGS (BNL) and JLab with CLAS (6\,GeV era).

The LEGS proton measurement spans $0.2\,\mathrm{GeV}  \leq \nu \leq 0.42\,\mathrm{GeV}$
and yields the contribution from single $\pi^0$ exclusive production $\gamma p\rightarrow \pi^0p$ to the integral of $(125.4 \pm 1.7 \pm 4.0)\,\mu\mathrm{b}$ \cite{Hoblit:2008iy}.

For the proton, the MAMI measurement covers  $0.2\,\mathrm{GeV} \leq \nu \leq 0.8\,\mathrm{GeV}$ and yields 
a contribution of $(254 \pm 5 \pm 12)\,\mu\mathrm{b}$. The ELSA measurement covers $0.7\,\mathrm{GeV} \leq \nu \leq 2.9\,\mathrm{GeV}$ and yields a contribution of
$(48.3 \pm 2.5 \pm 2.1)\,\mu\mathrm{b}$ \cite{Ahrens:2001qt}.

The JLab 6 GeV CLAS experiments (E04-102~\cite{E04-102} and E06-013~\cite{E06-013}, both part of the CLAS g9 run group) 
had as one of their goals to
measure some of the important photoproduction channels
contributing to the proton GDH sum. 
E04-102 measured the single-pion production contribution for $\nu$ up to 2 GeV, and
E06-013 measured the $\pi^+\pi^-$ contribution for $\nu$ up to 3.1 GeV.
These data are still under analysis and limited in their $\nu$ to the same
range as the MAMI and ELSA experiments. 
During the CLAS g14 run, which used the HDice target,
both proton and deuteron data were gathered, with part of the run using circularly polarized photons with $\nu$ up to 2.5 GeV. The goal of g14, however, was searching for missing resonances and the trigger 
was not suited for total cross-section measurements. Thus, we do not expect any direct information on the GDH sum from g14 and, in any case, its maximum $\nu$ coverage does not extend beyond that of ELSA. 
Finally, the 6 GeV CLAS experiment E94-117~\cite{GDH HB} was approved with an A$^-$ rating but did not run due to a delay in the polarized HDice target availability and the termination of the 6\,GeV program. 

Further GDH data on the proton and neutron have been acquired at GRAAL at ESRF (Grenoble, France, 0.5-1.5 GeV) and HIGS at TUNL (Durham, USA up to 0.1 GeV). However they are not published yet.  Another GDH experiment on deuterium is approved to run at HIGS in 2021, with beam energies between 6\,MeV and 20\,MeV.

An experiment with similar goals as this proposal, E159~\cite{E159}, was approved at SLAC but did not occur due to termination of the experimental program in End Station A. 

The GDH sum rule generalized for electroproduction has been the object of active experimental programs at ESRF, JLAB, MAMI, SLAC and TUNL, see Refs.~\cite{Helbing:2006zp, Deur:2018roz} for reviews.
A very low-$Q^2$ GDH program has been carried out in Halls A (E97-110~\cite{Sulkosky:2019zmn}) and B (E03-006 and E06-017~\cite{Adhikari:2017wox}) during the 6 GeV era. 
These experiments measured the inclusive doubly-polarized electron-scattering cross-section on proton and deuteron (eg4) and $^3$He and neutron (E97-110). 
The results can be extrapolated to $Q^2=0$ to investigate the GDH sum rule. However, due to elastic radiative tails rising at large $\nu$, the maximum $\nu$ value of these experiments was limited to 1.7 GeV at the lowest $Q^2$ used for the $Q^2 \to 0$ extrapolation. Therefore a low-$Q^2$ GDH program does not investigate the questions discussed in this proposal.

To summarize, the proton data are limited to about 3 GeV and the neutron data to 1.8 GeV. The LEGS, MAMI and ELSA $\Delta \sigma$ data are published, while the GRAAL, HIGS and CLAS data on specific channels are yet to be published.

\subsection{Current experimental status \label{Experimental Status}}
\subsubsection{Status of the GDH sum rule on the proton}

\begin{figure}[ht]
\centering
\includegraphics[width=0.75\textwidth]{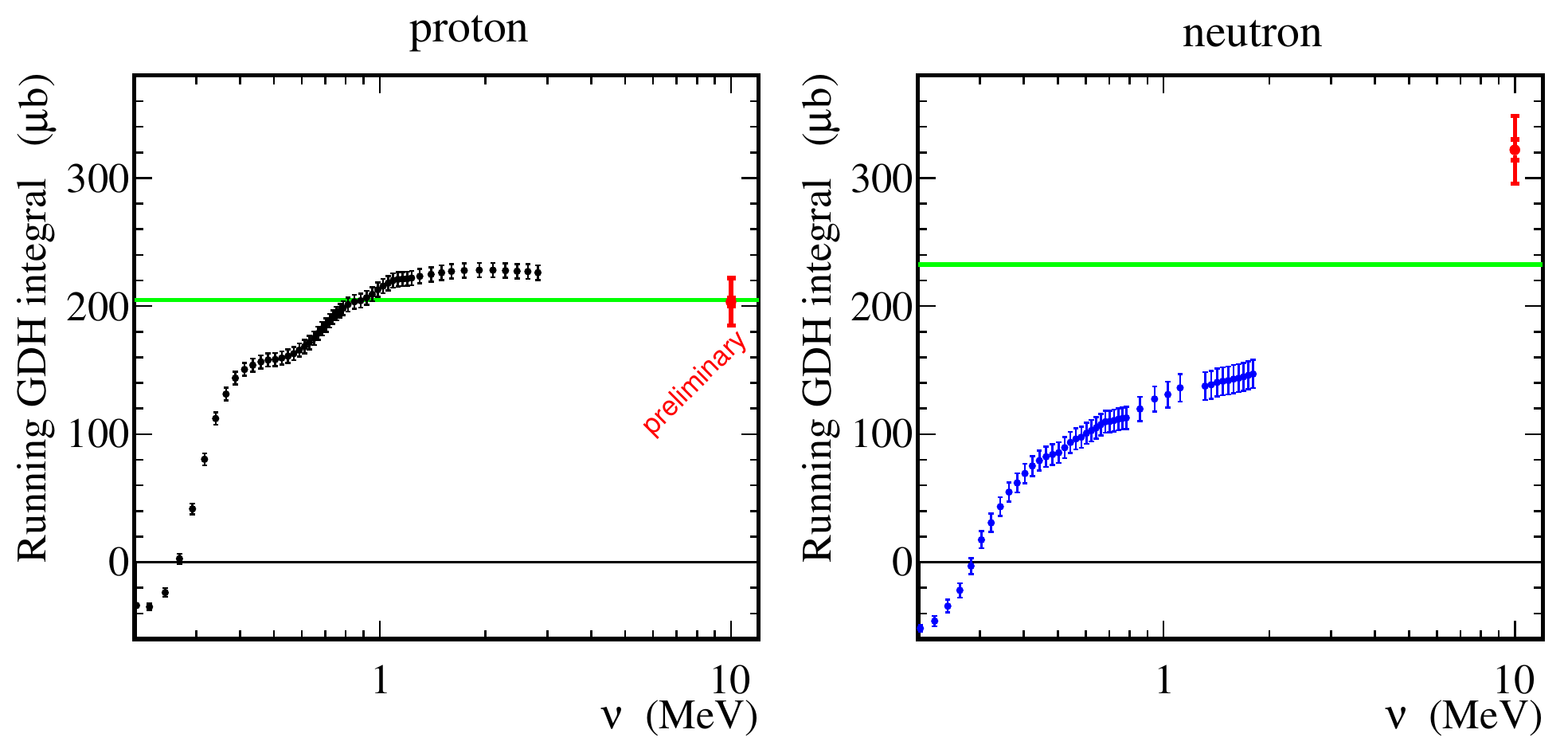}
\vspace{-0.3cm}
\caption{\label{fig:GDHrun_world}\small{``Running" of the GDH integral data for the proton (left) and neutron (right)
starting at $\nu=0.2$\,GeV, which is the smallest value 
so far measured.  
The $\nu_0 \le \nu\le 0.2\,\mathrm{GeV}$ contributions are estimated to be
$(-28.5 \pm 2)\,\mu\mathrm{b}$ and $\approx -41\,\mu\mathrm{b}$,
respectively, by the MAID2007 parameterization \cite{Drechsel:1998hk}.
The data from various photoabsorption experiments have been combined and rebinned.  The green horizontal lines show the expected value of the GDH sum.  The red points are the recent generalized GDH results from electroproduction extrapolated to $Q^2=0$ for the proton (preliminary, publication in preparation) and at $Q^2=0.035$ GeV$^2$ for the neutron~\cite{Sulkosky:2019zmn}, statistical uncertainty (inner) and total (outer) error bars.}}
\end{figure}

For the proton, the contribution from $\nu=\nu_0\approx 0.145\,\mathrm{GeV}$ 
to 0.2\,GeV  (from threshold to the start of the MAMI measurement) 
is estimated at $(-28.5\pm 2)\,\mu\mathrm{b}$ by the MAID2007 parameterization~\cite{Drechsel:1998hk}.  Including the measurements from Fig.~\ref{fig:GDH_world} and the MAID prediction yields an integral over the range $\nu=\nu_0$ to 2.9\,GeV of $(226\pm5.7\pm12)$\,$\mu$b.  Fig.~\ref{fig:GDHrun_world} shows the measured running of the GDH integral. To obtain the full integral, the unmeasured $\nu>2.9$~GeV contribution has been estimated using a Regge behavior, which had been argued to be adequate down to $\nu \approx 1.2$\,GeV for the spin-independent total photoabsorption cross-section~\cite{Helbing:2006zp}.
However, the appearance of the ``fourth resonance region" in the spin-dependent cross-section makes such an argument questionable.

\begin{figure}[ht]
\centering
\includegraphics[width=0.42\textwidth]{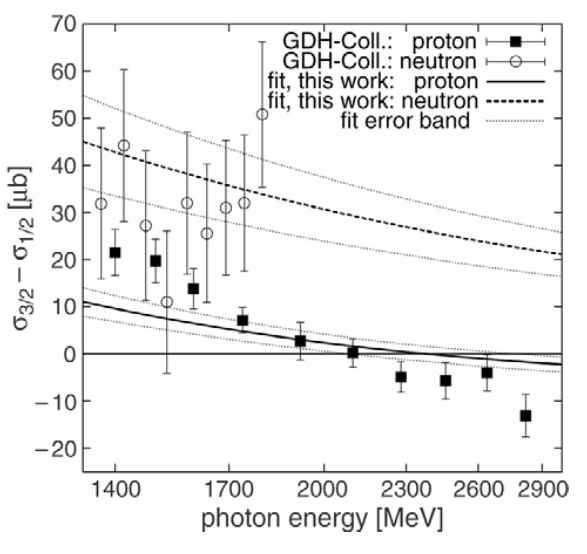}
\vspace{-0.3cm}
\caption{\label{fig:ReggeFail}\small{Simultaneous fit of Regge parametrization, Eq.~(\ref{eq:Regge_expect}), to existing spin-dependent data on proton and neutron.
Figure from Ref.~\cite{Helbing:2006zp}
}}
\end{figure}

Nevertheless, a Regge parameterization is assumed for the $\nu > 2.9$\,GeV contribution to the proton sum.  For fits done only on the proton data, contributions from $-20\,\mu\mathrm{b}$ to $-35\,\mu\mathrm{b}$ \cite{Helbing:2003pv} are obtained, 
where the range stems from the uncertainty on the $a_1$ intercept (parameter $\alpha^{a_1}$ in Eq.~(\ref{eq:Regge_expect})).
For combined fits on the proton and neutron data, the high-$\nu$ contribution is $-14\,\mu\mathrm{b}$~\cite{Helbing:2006zp} 
but the fit does not agree well with the proton data, 
as seen in Fig.~\ref{fig:ReggeFail}.
This suggests that a significant systematic uncertainty, which is difficult to quantify, is involved in the high-$\nu$ estimate  due to the limited range and quality of the existing data.
These projections show the full GDH integral lying in the range from 191\,$\mu$b to 212\,$\mu$b, which brackets the expected value by significantly more than the statistical uncertainty.

The CLAS eg4 electroproduction data extrapolated to $Q^2=0$ yield a preliminary\footnote{It is expected to be released in summer 2020.} result of $I^p=(203 \pm 19)\,\mu\mathrm{b}$, with a maximum energy coverage up to $\nu \approx 2.2$ GeV. The missing high-$\nu$ part is estimated by a parameterization of spin structure functions $g_1(Q^2, \nu)$ and $g_2(Q^2, \nu)$, including a Regge-based constraint for the highest $\nu$~\cite{Adhikari:2017wox}.

To summarize, the best estimates are compatible with the proton GDH prediction of $I^p=204.8$ $\mu$b with a 10\% accuracy, this one being dominated by the large-$\nu$ extrapolation, whose form is assumed to obey Regge theory, without it being verified for polarized photoabsorption. 
In fact, as seen in Fig.~\ref{fig:ReggeFail}, the extrapolation fits do not describe the data well. This important shortcoming of the current consensus on the status GDH sum rule can be addressed by the proposed GDH experiment in JLab Hall D.

\subsubsection{Status of the GDH sum rule on the neutron}

No assessment of the neutron GDH sum rule has been published yet.
We formed the neutron GDH running integral by using the published world data and show it in Fig.~\ref{fig:GDHrun_world}. The MAID model is used to estimate the unmeasured low-$\nu$ contribution. Also shown is the GDH integral  generalized to electroproduction, measured at $Q^2=0.035$~GeV$^2$~\cite{Sulkosky:2019zmn} and including an estimate of the high-$\nu$ part. How the generalized integral evolves to $Q^2=0$ is an unsettled question (the state-of-art $\chi$EFT estimates disagree). However, at such low $Q^2$, the evolution to $Q^2=0$ is expected to cause only a small change. Regardless of this question, Fig.~\ref{fig:GDHrun_world} illustrates the present lack of convergence and validity check of the neutron GDH integral.

\subsection{Primary goals and proposal layout}

This proposal will utilize the polarized 12\,GeV CEBAF beam to measure the  high-energy behavior of the GDH sum on the proton and neutron, with two primary objectives:
\begin{itemize}
    \item The \textit{convergence} of the GDH integral will be studied through a measurement of the yield difference $\Delta y(\nu) \propto \Delta \sigma (\nu)$. This eliminates uncertainties coming from  normalization factors and  unpolarized backgrounds. 
    \item The \textit{validity} of the GDH sum rule will be studied through a measurement of $\Delta \sigma (\nu)$, which will more extensively test Regge and chiral effective field theories. 
\end{itemize}

Hall D, with its high-luminosity photon tagger and its large solid angle detector is  uniquely suited for such an experiment.
The experiment is not feasible in other JLab Halls since they lack the photon tagging capability of Hall D and a GDH measurement via electroproduction is poorly matched to a study of its large $\nu$ domain. The measurement would have to be done at low enough $Q^2$ so that a reliable extrapolation to $Q^2=0$ could be done. 
However, with an 11 GeV beam, this would require a measurement at scattering angles smaller than 0.8$^\circ$, which no Hall can reach and where the elastic radiative tails are prohibitively large. 

The Hall D measurement would extend by a factor of 4 the experimental integration range for the proton and by a factor of 7 for the neutron. 
It will cover the domain relevant to clarify the question of the convergence of the GDH integral and the validity of Regge theory for the nucleon spin structure, while probing  for unknown parton process or structure.

The proposal is laid out as follows.  
In Section~\ref{sec:theory} we sketch the derivation of the sum rule, discuss the no-subtraction hypothesis and discuss potential mechanisms that may cause the sum rule to be violated.
In Section~\ref{sec:experiment} we examine the experimental requirements needed for a measurement of the spin-dependent cross-section on the proton and neutron.
In Section~\ref{sec:simulation} we describe the simulation of the experiment signal and backgrounds.
In Section~\ref{sec:uncert_stats} we discuss the expected statistical uncertainties and the implications for the  analysis of the functional forms.
In Section~\ref{sec:uncert_syst} we discuss the systematic uncertainty expected on the absolute cross-section given conservative assumptions.
In Section~\ref{sec:request} we summarize the total time requested for the experiment.
In Section~\ref{sec:impact} we discuss the impact of the experiment on Regge phenomenology, the spin-dependent Compton amplitude, and diffractive physics.

\section{Theory}\label{sec:theory}

\subsection{Derivation of the GDH sum rule \label{GDH derivation}}

Several methods have been used to derive the GDH sum rule
\cite{Helbing:2006zp, Pantforder:1998nb}.  To elucidate what
the sum  rule actually tests, we outline here the derivation using 
the dispersion relation approach.  It starts from the forward real 
Compton scattering amplitude $F(\nu)$ and utilizes causality 
(dispersion relation); unitarity; Lorentz and gauge invariances 
(low energy theorem).   While studying the GDH sum rule tests 
all these hypotheses, the latter two are robust and stand 
at the foundation of quantum field theory.  The exception, 
the ``no-subtraction hypothesis", enters the derivation 
of the first item, the dispersion relation.  Its validity depends
on the observable involved: for a nucleon target, it involves QCD 
in general and specifically the high-energy behavior of $F(\nu)$.  There has been much discussion on whether 
the no-subtraction hypothesis holds in the context of the nucleon 
GDH sum rule; see, for instance, Ref.~\cite{Bass:1997av}. 

The forward Compton amplitude $F(\nu)$  depends on the polarization of the incoming and scattered
photons, $\pmb{\epsilon_1}$ and $\pmb{\epsilon_2}$, respectively, 
and on their momenta which, for forward scattering, obey $\pmb{k_1} = \pmb{k_2} \equiv \pmb{k}$. 
Five functions $f_i(\nu)$ can then be defined to parameterize $F(\nu)$ 
since it is a scalar quantity.  For real photons $\pmb{k\cdot\epsilon}=0$, 
which reduces the number of parameters to two:
\begin{equation}
F(\nu) = f_1(\nu)\,\pmb{\epsilon_2}^*\cdot\pmb{\epsilon_1} + 
f_2(\nu)\,\pmb{\sigma} (\pmb{\epsilon_2}^* \times \pmb{\epsilon_1}) \>,
\end{equation}
where $\pmb{\sigma}$ are the Pauli matrices. The spin-independent 
amplitude $f_1(\nu)$ is used in the (unpolarized) Baldin sum rule 
derivation \cite{baldinSR}, while the spin-dependent amplitude
$f_2(\nu)$ yields the GDH sum rule.
Causality implies the analyticity of $f_2(\nu)$ in the complex plane, 
yielding the Cauchy relation:
\begin{equation}
f_2(\nu)
  =\frac{1}{2i\pi }\oint \frac{f_2(\varepsilon)}{\varepsilon-\nu}
   \,\mathrm{d}\varepsilon
  =\frac{1}{2i\pi }\int ^{+\infty }_{-\infty }\frac{f_2(\varepsilon)}
    {\varepsilon-\nu} \, \mathrm{d}\varepsilon \>.
\label{eq:Cauchy}
\end{equation}
The right-hand side equality holds if the Jordan lemmas are valid 
for $f_2(\nu)$, that is, if $ f_2(\nu)$ vanishes when $\nu\rightarrow\infty$.
In that case
\begin{equation}
\Re e \big(f_2(\nu ) \big)=\frac{1}{\pi }P\int ^{+\infty }_{-\infty }\frac{\Im m \big(f_2(\varepsilon) \big)}{\varepsilon-\nu }\,\mathrm{d}\varepsilon\>,
\label{eq:KK}
\end{equation}
which is the Kramer-Kr\"{o}nig relation~\cite{KKR}, ubiquitous 
to all fields of physics. 
Crossing symmetry implies $f_2(\varepsilon )=-f_2(-\varepsilon )^*$ which, 
applied to Eq.~(\ref{eq:KK}), yields
\begin{equation}
\Re e \big(f_2(\nu )  \big)=\frac{2\nu }{\pi }P\int ^{+\infty }_{0}
\frac{\Im m  \big(f_2(\varepsilon)  \big)}{\varepsilon^{2}-\nu ^{2}}\,
\mathrm{d}\varepsilon \>.
\label{disprel}
 \end{equation}
Unitarity gives
\begin{equation}
\Im m \big(f_2(\varepsilon )\big)=\frac{\varepsilon }{8\pi }(\sigma_A-\sigma_P) \>.
\label{eq:optheo}
\end{equation}
A low energy theorem (Lorentz and gauge invariances, and crossing symmetry) 
can be used to expand $f_2$ in $\nu$: 
\begin{equation}
f_2(\nu )=-\frac{\alpha \kappa ^{2}}{2M^2}\nu 
  +\gamma \nu ^{3}+{\cal O}(\nu ^{5}) \>.
\label{eq:LET}
\end{equation}
The derivative of Eq.~(\ref{eq:LET}) together with 
Eqs.~(\ref{eq:optheo}) and (\ref{disprel}) yield
the GDH sum rule:
\[
\left.\frac{df_2(\nu )}{d\nu }\right| _{\nu=0}
  = \frac{\alpha \kappa ^{2}}{2M^{2}}
  = \frac{1}{4\pi ^{2}}\int ^{\infty}_{\nu_0}
    (\sigma _P-\sigma_A)\frac{\mathrm{d}\nu}{\nu}
\]
where we have changed the dummy variable $\varepsilon$ to $\nu$ in the integral. 

\subsection{Pole and subtraction hypothesis\label{subhyp}}

One of the mechanisms that could compromise the above
derivation and lead to a violation of the GDH sum rule is 
the possibility of a $J=1$ pole of the Compton amplitude
\cite{Abarbanel:1967wk}.  Such a pole would invalidate the Jordan 
lemma since $\Re e(f_2)$ would not vanish as $\nu \to \infty$,
$\Re e\big(f_2(\infty)\big) \neq 0$.  But $\Im m(f_2)$ would still 
vanish and thus a pole would not affect the overall convergence 
property of the GDH integral $I$.  It would, however, affect 
the $\nu-$dependence of $\Delta \sigma(\nu)$ since it would add
a constant to the GDH relation coming from the contribution 
of circle integration that was assumed to vanish in the right-hand 
side of Eq.~(\ref{eq:Cauchy}).  This would lead to 
a ``subtracted GDH sum rule''
\begin{equation}
I \equiv \int_{\nu_0}^{\infty}\frac{\Delta \sigma(\nu)}{\nu}\,\mathrm{d}\nu
  =\frac{2\pi^2 \alpha\kappa^2}{M^2} -4\pi^2\Re e\big(f_2(\infty)\big) \>.
\label{eq:gdh+pole}
\end{equation}

As discussed in~\cite{Pantforder:1998nb}, a pole would be related 
to the behavior of a Compton amplitude at high energy. 
In fact, the current data also indicate that if a pole is present, 
it would manifest in the high-$\nu$ behavior of $\Delta \sigma$: 
the data show that the resonance region saturates the GDH sum,
$$
\int_{\nu_0}^{\approx 2~\mbox{{\scriptsize{GeV}}}}\frac{\Delta \sigma(\nu)}{\nu}
  \,\mathrm{d}\nu \approx \frac{2\pi^2 \alpha\kappa^2}{M^2} \>.
$$
Thus, the additional term $-4\pi^2\Re e\big(f_2(\infty)\big)$ in Eq.~(\ref{eq:gdh+pole}) must come from the behavior of $\Delta \sigma$ 
at higher $\nu$.

\subsection{Possible causes of violation \label{sec:violation}}

Possible causes for a GDH sum rule violation---or its apparent violation when the integral is measured over a finite $\nu$-range---are reviewed in~\cite{Pantforder:1998nb}.
The ones most often considered are 
a) the existence of unknown high-energy phenomena, such as quark substructure (non-zero quark anomalous moments)~\cite{Kawarabayashi:1966irp}. 
b) The existence of a $J=1$ pole of the nucleon Compton amplitude~\cite{Abarbanel:1967wk} as just discussed in Section~\ref{subhyp};
and c) the chiral anomaly~\cite{Chang:1994td}.
All proposed mechanisms would manifest themselves at high $\nu$.
Since there is no low-$\nu$ mechanism that could invalidate the sum rule, and since the convergence can be investigated only beyond the resonance region, to truly verify the sum rule, the behavior of $\Delta \sigma$ at high $\nu$ must be measured.

\section{Experiment \label{sec:experiment}}

Testing the \textit{convergence} of the GDH integral requires only the shape of the high energy part of the integrand. 
It therefore suffices to  measure the yield difference $\Delta y(\nu) = N^+ - N^-$, where $N^{+(-)}$ is the number of events in a bin $\nu$ for positive (negative) beam helicity.
This quantity is insensitive to normalization uncertainties which are typically dominant in  
experiments measuring cross-sections.  
Furthermore, uncertainties from the unpolarized contributions (target dilution) cancel in the $N^+ - N^-$ difference.
For the integral to converge, $| \Delta \sigma / \nu |$ must decrease with $\nu$ (baring exotic behavior such as a singular contribution at $\nu\to \infty$), and thus only 
the $\nu$-dependence of $\Delta \sigma$ must be established 
in order to assess the convergence.
Recall that such a decrease does not occur for $| \sigma|$ and that the unpolarized equivalent of the  GDH sum does not converge.  

Testing the \textit{validity} of the sum rule will require 
normalizing $\Delta y(\nu)$ into a cross-section by measuring the beam flux,  target density, solid angle, target and beam polarisation, as well as detection efficiencies.

The signal of interest is the total spin-dependent yield of photoproduced hadrons, that is simply counting events with at least one hadron in the final state and a reaction invariant mass greater than the nucleon rest mass. The three main ingredients needed for measuring the spin-dependent yield are:
\begin{itemize}
	\item a beam of circularly polarized tagged photons;
	\item a longitudinally polarized target; 
	\item a large solid-angle detector. 
\end{itemize}
%

\subsection{Beam}

\subsubsection{Photon-beam production and polarization}\label{sec:photonbeam}

\begin{figure}[ht]
\begin{center}
\vspace{-0.cm}
\includegraphics[width=7.7cm]{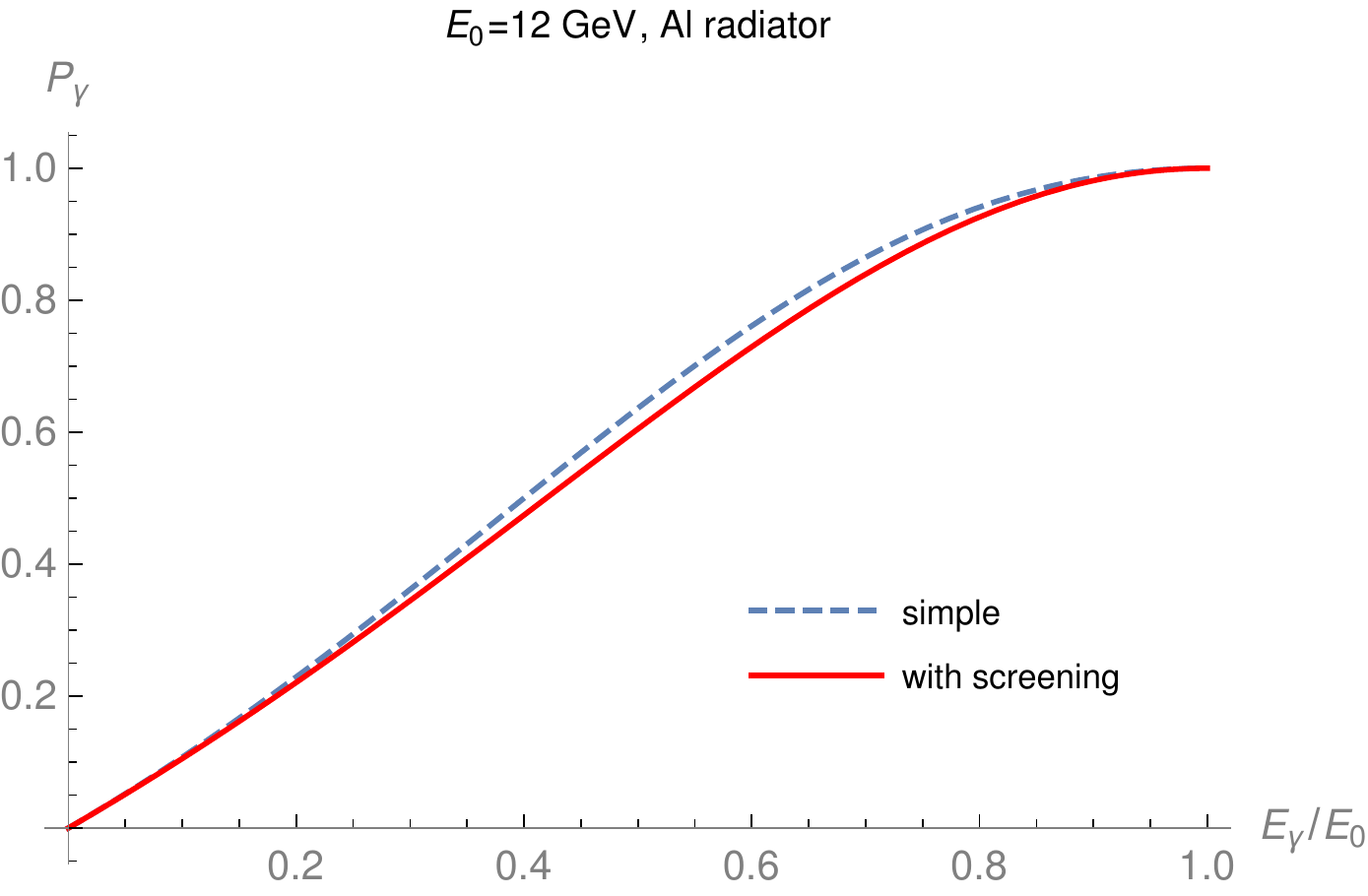}
\includegraphics[width=7.7cm]{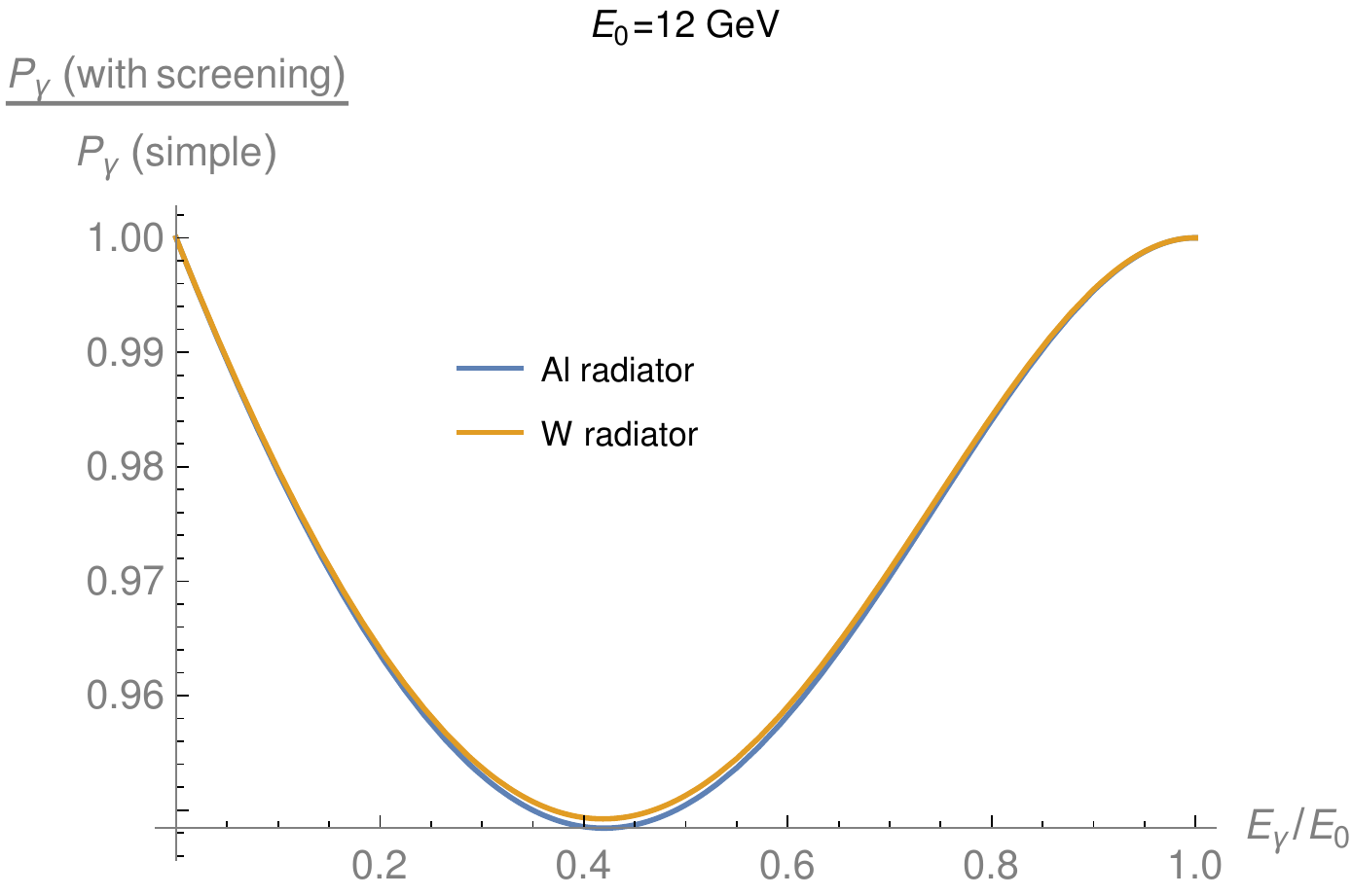}
\end{center}
\vspace{-0.3cm}
\caption{\label{Flo:circpol}\small{Left: photon circular polarization versus the energy fraction $\nu/E$ using the approximation given by
Eq.~(\ref{photon polarization}) (dashed blue curve), and the exact formula 
for aluminum (full red curve).  Right: ratio between the exact  
and approximate calculations for different radiator materials 
(blue: aluminum; orange: tungsten).
The results here assumes 100\% electron beam polarization and the curves on the left panel need to be rescaled by the actual electron beam polarization, assumed to be 80\% in this document. }}
\end{figure}

Circularly polarized photons are necessary to measure $\sigma_{P}$ 
and $\sigma_{A}$.  They can be generated using CEBAF's polarized 
electrons with an amorphous radiator.
Their polarization can be approximated by \cite{Olsen:1959zz}:
\begin{equation}
P_\gamma \approx P_e\frac{y(4-y)}{4-4y+3y^2} \>,
\label{photon polarization}
\end{equation}
where $y = \nu/E$, $E$ is the electron beam energy and $P_e$ is the electron beam polarization. 
$P_\gamma(y)$ obtained by the 
approximation~(\ref{photon polarization}) and by the exact 
formula are shown on Fig.~\ref{Flo:circpol}. Also shown 
is the effect of using different radiator materials for the exact formula. The material of the radiator is of little importance from a polarization point of view.

In this proposal $P_e$ is assumed to be 80\%.
In terms of the figure-of-merit, the increase of photon beam polarization 
at higher $\nu$ more than compensates the decreasing flux and cross-section. 
Thus, we expect a better statistical precision at larger $\nu$, 
see Figs.~\ref{fig:expect} and \ref{fig:expect_n} for the projected results.
 
No electron beam polarimetry is presently available in Hall D.%
\footnote{It is relatively easy and of moderate 
cost to build a M\"oller  polarimeter for the Hall D beam.
However, this may not be warranted solely for the purpose 
of this single experiment. Although we anticipate that this 
experiment will initiate a program using circularly polarized 
photons in Hall~D, we conservatively assume in this document 
that there will be no polarimeter available.}
The electron beam longitudinal polarization can be measured 
at $<1$\% level using the injector Mott polarimeter or the polarimeters in Halls A or C. Spin precession can be calculated to few degree accuracy for a beam energy known at the $10^{-3}$ level, the presently known accuracy on the beam energy in Hall D.  In order to bound the time variation of the polarization, Mott polarization measurements at the source will be necessary, if no polarization measurement is done in the other Halls. 
Calculations indicate that the depolarization resulting from 
synchrotron radiation and energy spread are below 1\%, which 
is confirmed by the high beam polarizations measured in Hall~A and B at 11\,GeV \cite{J.Grames}.
 
There is presently no equipment in Hall D to monitor, record and control the electron beam helicity information and its charge asymmetry. Its implementation is straightforward and its cost estimated to be less than \$50k.

\subsubsection{Photon flux and energy measurement}
\label{sec:photonmeasure}

The photon flux is monitored by the Hall D Pair Spectrometer (PS), which has been calibrated at the percent level by dedicated runs performed at very low current:  A calorimeter is inserted into the beam to measure the flux by counting every photon.  Two such devices are available, the Total Absorption Counter (TAC) and the Compton Calorimeter (CCAL).  This is more than sufficient for the present proposal. 

As it exists, the PS covers a momentum range of about $\pm30\%$.  It would thus be necessary to run the PS at three different fields during the experiment to correct the flux for the tagger inefficiency and collimator transmission over the full $\nu$-range of the experiment.  The energy ranges would be approximately 3.0\,GeV to 5.6\,GeV, 4.2\,GeV to 7.7\,GeV and 6.6\,GeV to 12.0\,GeV. 
Alternatively, one could upgrade the PS detectors so that they cover the low energy photon flux by adding detector paddles at larger angles.
A possibly more attractive option would be to move the PS detectors closer to the PS magnet so that they cover a larger energy bite, with an associated decrease in the energy resolution of the pair. 

The photon energy is tagged with a resolution better than 0.5\% \cite{ASomov} which is more than sufficient for the present proposal.

\subsubsection{Helicity correlated beam asymmetries }

As with any experiment that measures a beam spin asymmetry at CEBAF, it is necessary to quantify potential false asymmetries that arise from beam properties that change with the polarization orientation, i.~e.~helicity-correlated beam asymmetries.
The experiment is insensitive to angle and position differences at the target as it is completely azimuthally symmetric.  It is, however, sensitive to potential beam motion at the collimator which may change the transmitted flux.  Simulations were done of the transmission of the beam through the collimator for two beam sizes which bound the usual size of the beam, 0.5 mm and 1 mm at the collimator.  Larger transverse beam size has lower transmission but less sensitivity to beam motion.  It was found that for a photon beam from an amorphous radiator (bremsstrahlung without a coherent component), both the transmission and the sensitivity to beam motion are independent of the photon energy.

For a beam size of 0.5 mm at the collimator, the transmission is well modeled by $T = 0.25 - 0.016\,x^2$ where $x$ is the offset of the beam from the center of the collimator in mm.

The 12 GeV CEBAF Beam Parameter Tables\footnote{\href{https://www.jlab.org/physics/PAC}{12 GeV CEBAF Beam Parameter Tables}} indicate that we can achieve helicity correlated positions at the target $<25$\,nm averaged over an 8 hour period.  A 25\,nm position difference at the Hall D collimator would lead to a $10^{-11}$ relative change in rate if the beam is centered on the collimator and a $3\times10^{-6}$ relative change in rate if the beam is 1~mm off center.
This is negligible compared to the size of the trigger asymmetry in the experiment, about $2\times10^{-3}$ when dilution from the full butanol molecule is considered.  We expect that no correction will be made,  nevertheless it would be prudent to measure the electron beam position in the tagger hall with the DAQ.  Values much smaller than 25\,nm are routinely achieved for parity violation experiments.  In practice, we could tolerate position differences up to 250\,nm if the beam is centered on the collimator.

\subsection{Target \label{target}}

For this experiment, we propose to design and construct
a frozen-spin polarized target that can serve 
as the foundation for a new polarized target program in Hall D.  Protons and deuterons in samples of butanol and d-butanol will
be dynamically polarized outside the detector in fields up
to 5~T and temperatures around 0.3~K.  Proton polarizations
above 90\% and deuteron polarizations approaching
90\% have been previously demonstrated in a similar
system in Hall B.
Once polarized, the
sample temperature is reduced below 50~mK, and the
target is retracted from the polarizing magnet
and moved into the Hall D detector magnet.  A thin, 0.5~T
superconducting solenoid will be incorporated inside the
target cryostat to maintain the polarization while in transit.
Data is acquired while the nuclear polarization slowly
decays in an exponential fashion characterized by the
$1/e$ time constant $T_1$. In-beam values of $T_1$ of 2800~h
(1400~h) 
were obtained, with the polarization parallel (anti-parallel) to the holding magnetic field, in Hall B using a 0.56~T holding field---meaning
a polarization loss of 2\% or less per day.  The Hall B
experiments were halted about once per week to reverse the
polarization, which required 4--8 hours.  The greater
photon flux envisioned for experiments in Hall D will produce
a warmer sample temperature and decrease $T_1$, but this
will be offset by the higher holding field
of the Hall D magnet (1.8~T).  For this proposal we assume
an average polarization of 80\%,\footnote{An average of 82\% was achieved during the Hall B g9a run.}
with a 3\% accuracy on the polarimetry.

The target sample will be 7~mm in diameter and 
100~mm long, and for optimum cooling will be comprised of multiple 1--2~mm beads of frozen butanol (C$_4$H$_9$OH) 
or its fully deuterated counterpart.  These
will be chemically doped with an appropriate paramagnetic
radical for dynamic nuclear polarization, the nitroxyl radical
TEMPO in the case of protons and the
trityl radical CT-03 for deuterons. 
Assuming a 60\% packing fraction for the beads, the target density will be about 0.66 g/cm$^3$ for butanol and 0.73 g/cm$^3$ for d-butanol.

A $^{12}$C foil will be placed upstream of the polarized sample 
to allow the extraction of the relative asymmetry ${\Delta \sigma}/{\sigma}$ and is needed to correct the asymmetry for the dilution  by unpolarized target material.%
\footnote{The carbon foil data will allow to compute the number of counts $N^0$ from the unpolarized part of the butanol, i.~e.~the dilution $D$ of the asymmetry. Thus, by measuring the diluted asymmetry $(N^+ - N^-)/(N^+ + N^- +2N^0) \equiv \Delta \sigma / D\sigma$ with the butanol target, and $D$ with the carbon foil, we obtain the physics asymmetry $\Delta \sigma / \sigma$.}   Although not needed to achieve the goals of this proposal, forming this asymmetry will permit a fast initial 
analysis and offer a thorough verification of the main analysis method.\footnote{Asymmetry-based analyses are generally easier and faster than absolute cross-section analyses, and often more accurate. However, in the context of this experiment, the cross-section difference method is more accurate. In addition, asymmetry analyses do not provide information on the absolute cross-section, this one being assumed to be available from world data.}
Since the diluting material is mostly carbon and oxygen from the butanol, the dilution is essentially obtained by scaling the $^{12}$C foil rate by (4 + 16/12), correcting it for a small detector acceptance effect, and dividing it by the butanol rate.  This factor is then used to correct the raw asymmetry.   
 A 3~mm $^{12}$C foil
(0.35~g/cm$^2$, or 5\% of the polarized target thickness) 
should be adequate.  As was done in Hall B,
the foil can be mounted on a heat shield a few cm 
downstream of the butanol sample for accurate separation based on vertex reconstruction.
Dilution from the $^3$He--$^4$He bath and beam line windows is determined separately, but in the same way, by empty target runs.

The anticipated total photon intensity on the target
is $7 \times 10^7\,\gamma/$s, leading to a
heat load from e$^+$e$^-$ pair production of
approximately 14~$\mu$W on the butanol sample.
The $^3$He-$^4$He dilution refrigerator built for the
Hall B frozen spin target could absorb this heat load and
maintain a temperature below 40~mK. 
However, the butanol beads will warm substantially
above this temperature due to the thermal boundary 
(Kapitza) resistance between the beads and the 
$^3$He-$^4$He coolant.  The average temperature of
the beads, assuming the heat load is equally shared 
among all beads in the beam path, can be written as
\begin{equation}
	T_b = \left( \frac{\dot q}{4 \pi r^2 \alpha} + T^4_c \right)^{1/4}\>.   \label{eqn:Thot}
\end{equation}
Here, $\dot q$ is the heat load on an individual bead of butanol
of average radius $r = 0.75$~mm, $T_c$ is the temperature of the coolant bath, and $\alpha$ is Kaptiza conduction 
between the bath and coolant.  The latter is difficult to
estimate, but we take a value 
$\alpha = 28$~W~m$^{-2}$~K$^{-4}$ based on a survey of
the available measurements\footnote{C.D.~Keith, {\em Polarized
Targets in Intense Beams\/}, unpublished.}  and
find a bead temperature of 0.18~K.

The relaxation time of protons in butanol in a 0.56~T
field has been measured to be $T_1 = 30$~hr at a slight lower
temperature of 0.15~K. 
Assuming $T_1 \propto B^3$, 
we estimate $T_1$ should be 900--1000~h in the 1.8~T field of
the Hall D detector.  This can be increased to more
than 2000~h if the 0.5~T transit coil remains energized
while in the detector.
Additionally, some of the beam heating can be alleviated 
using a photon beam hardener to reduce
the low energy ($<100$~MeV) portion of the 
flux~\cite{gamhard}. Such hardeners have been used at
SLAC, CEA and DESY to suppress low energy bremsstrahlung
photons.

Still higher fluxes can be sustained if we 
replace the 0.5~T transit coil with a slightly stronger coil that increases the net field inside the Hall D detector to 2.5--3~T and improves its field uniformity to a level suitable for dynamic polarization ($\sim$100~ppm).  Doing so would allow us to {\em continuously} polarize target samples inside the
detector and run with one or two orders of magnitude 
greater photon fluxes (maximum DAQ rate and tagger accidentals permitting).  In this case, radiation 
damage to the target material may become the limiting factor, 
but this can be countered by using more radiation resistant sample materials such
NH$_3$ and ND$_3$.

Based on previous experience, two months will be required
to install and test the target in Hall D.  Polarization of proton samples requires 4--8 hours, and up to 24 hours is needed to reach the maximum polarization of deuterons.  A similar
amount of time is required to reverse the polarization,
although we may pursue RF methods such as adiabatic fast passage
to speed this process if frequent reversals are desired.
Otherwise, these will be coordinated to coincide with
weekly beam studies and RF recoveries of the accelerator cavities.
While not absolutely necessary, taking beam on both
target polarization directions can reduce systematic  biases  in  the  
measured  cross-section  asymmetry.
Replacing one sample with another is an additional four hours.  

The estimated cost of this system is about \$600k, assuming
that components from the previous Hall B target are re-purposed.
A new, custom $^3$He-$^4$He dilution refrigerator will be needed to
accommodate the present cryogenic capabilities of the Hall D 4~K
refrigerator. This can be designed and constructed by the JLab
Target Group, who built the FROST refrigerator for Hall B.

\subsection{Detectors and rates}

\begin{figure}[hbt]\centering
\includegraphics[width=0.7\textwidth]{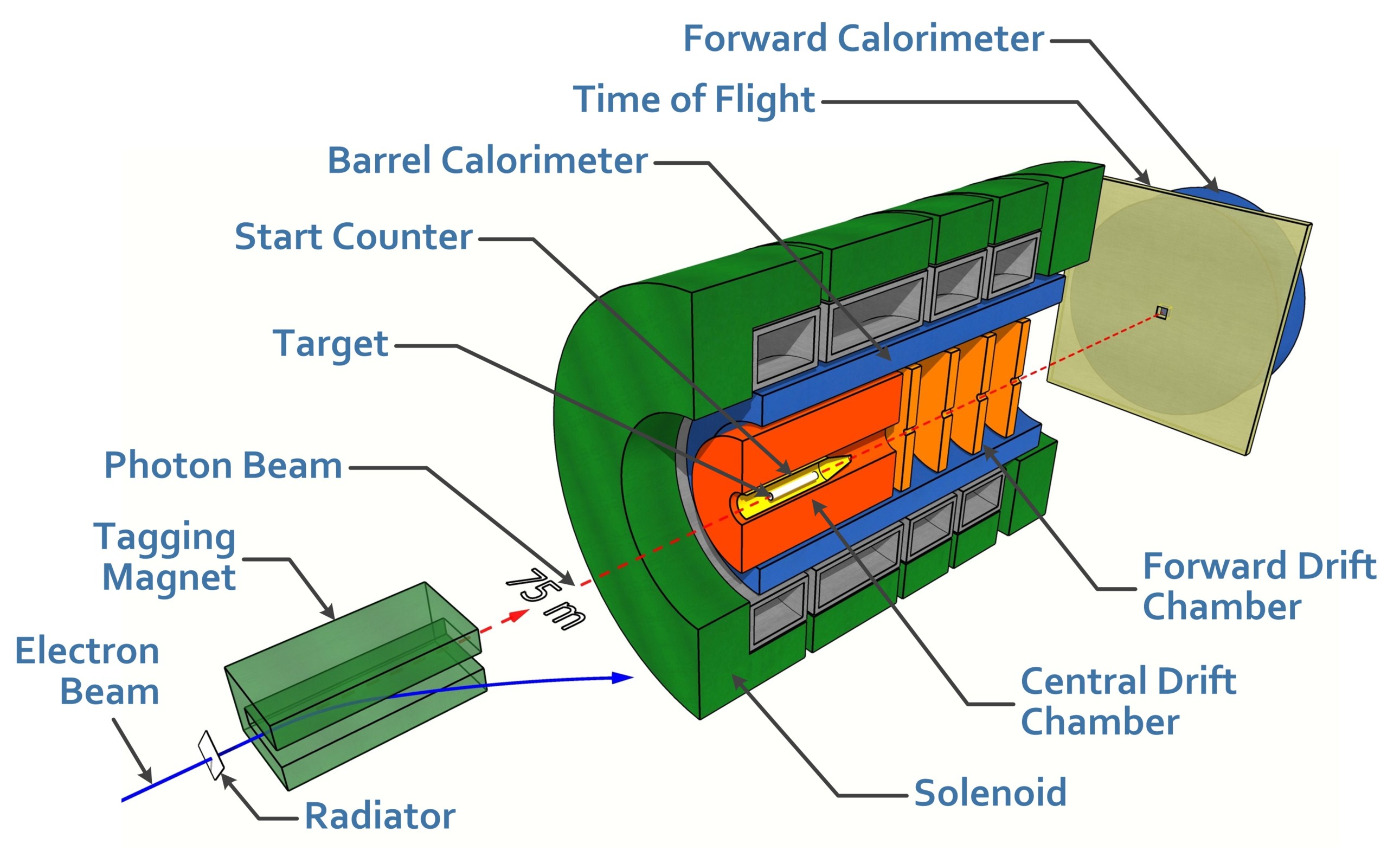}
\caption{\label{fig:gluex_cut-away}A cut-away drawing of the GlueX detector in Hall D from Ref.~\cite{Adhikari:2020cvz}, not to scale.}
\end{figure}

Hall D is uniquely suited at JLab to measure the total photoproduction cross-section thanks to its large solid angle and its tagger.
The standard GlueX/Hall D detector package plus the recently commissioned PrimEx-$\eta$ Compton Calorimeter is assumed in this proposal.  The GlueX beamline and detector are shown in Fig.~\ref{fig:gluex_cut-away} and described in detail in Ref.~\cite{Adhikari:2020cvz}, with the relevant components summarized here. 

\paragraph{Trigger}

The main trigger requirement is similar to that of GlueX, 
i.e.~set to accept most hadronic events while reducing the electromagnetic  background rate.  Hence, the main trigger will be similar to the GlueX one and based on measurement of energy depositions in  the   calorimeters (the Barrel Calorimeter, BCAL, and the Forward Calorimeter, FCAL). The GlueX trigger requires $E_\text{BCAL}+2E_\text{FCAL}>1$\,GeV but eliminates the three most inner blocks of the FCAL from the sum.

\paragraph{Drift Chambers}

Use of the existing drift chambers is required in order to do charged particle tracking and PID by $dE/dx$ as is currently done in GlueX.
For those forward-going particles that do not hit the Start Counter,  tracking is required in addition to the Time Of Flight (TOF) wall in order to provide precise timing that would identify the tagged photon responsible for the detected event.
Tracking information is required in order to do exclusive channel reconstruction---which is used for systematic studies of the detector acceptance and efficiency as well as the photon beam tagging.  Additionally, potentially ancillary results are possible for specific exclusive final states, as described in Sec.~\ref{sec:exclusiveanalysis}.

\paragraph{Calorimeters}

The calorimeters provide detection of neutral and charged particles over polar angles from 0.2$^\circ$ to 145$^\circ$ with a nearly complete azimuthal coverage.  The trigger relies only on energy distribution in the calorimeters, where the location and amount of the deposited energy required for a trigger may be tuned.
Neutral particles are detected by the Forward Calorimeter (FCAL) between 1$^\circ$ and 11$^\circ$ and the Barrel Calorimeter (BCAL) between 12$^\circ$ and 160$^\circ$. The Compton Calorimeter (CCAL) covers forward angles down to 0.2$^\circ$. 
Section~\ref{sec:analysis} shows that the acceptance for inclusive events is high and therefore the expected acceptance correction will be small.

\paragraph{Rates}

\begin{figure}[ht!]
\begin{center}
\vspace{-0.cm}
\includegraphics[width=0.8\textwidth]{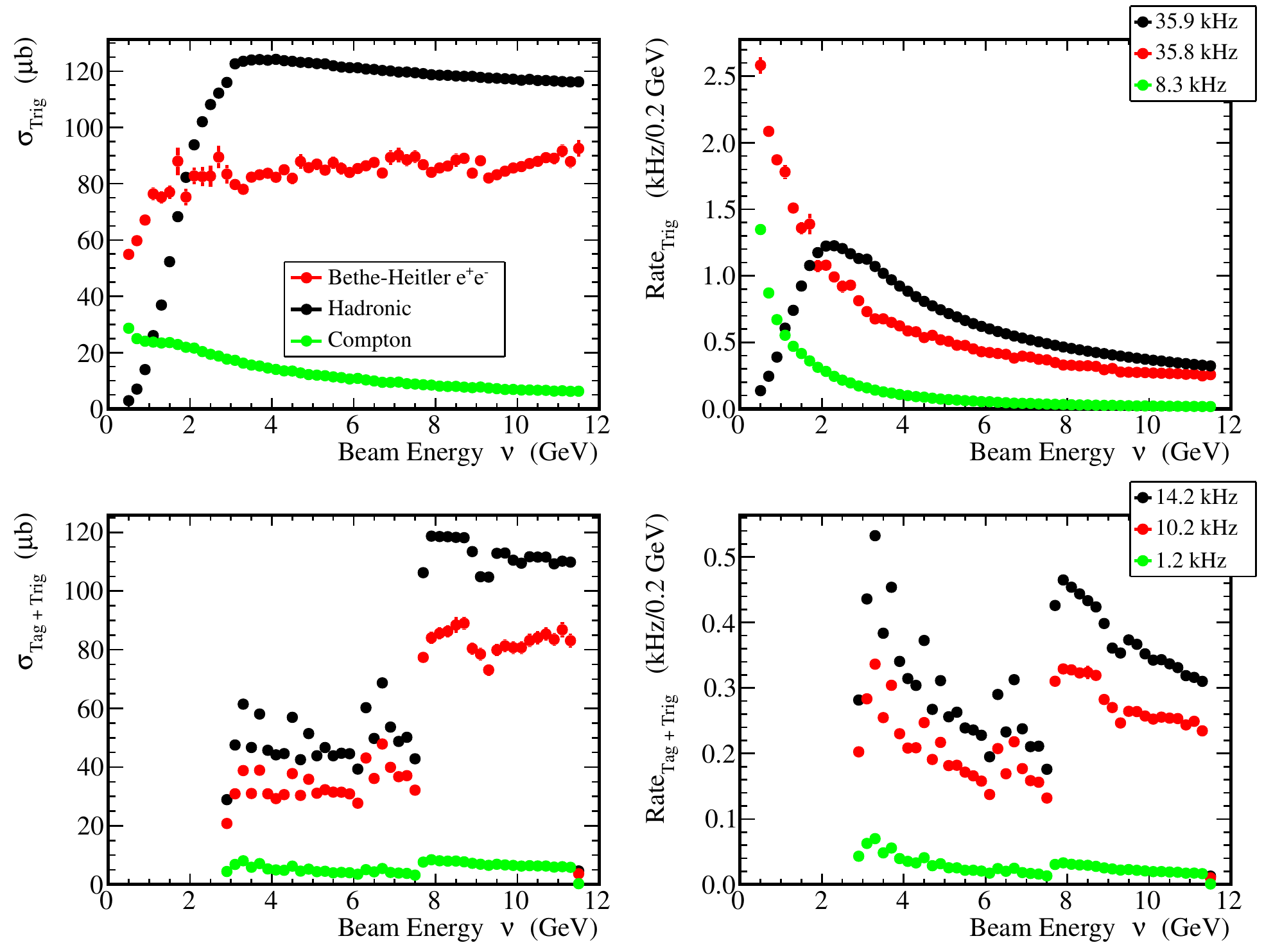}
\end{center}
\vspace{-0.3cm}
\caption{\label{Fig:XS and rates}\small{{Cross-sections (left) and rates (right) for the total hadronic photoproduction (black), the Bethe-Heitler background (red) and the Compton background (green).   
The top panels show the trigger cross-sections (total cross-sections multiplied by trigger efficiency) and rates using the standard GlueX trigger.
The flux is chosen so that the trigger rate equals the present maximum DAQ rate of 80 kHz.
The bottom panels show the cross-sections and rates after
accounting for the tagger acceptance and efficiency.}}}
\end{figure}

Imposing a 80\,kHz data acquisition limit, the 120 $\mu$b total $\gamma p$ unpolarized cross-section yields a total hadronic rate of 35.9 kHz%
\footnote{The dilution factor of butanol is about $10/74=0.135$ (proton) or $20/84=0.238$ (deuteron), and the Carbon foil is 5\% of the main target thickness, which yields a useful rate of 1.8 kHz and 3.2 kHz, respectively, for the proton and deuteron.},
see top right panel of Fig.~\ref{Fig:XS and rates}. 
The Bethe-Heitler and Compton background rates 
are 35.8 kHz and 8.3 kHz, respectively.
We neglected the small target wall and cosmic rate contributions. Section~\ref{sec:backgrounds} gives details on the treatment of the backgrounds.
Accounting for the tagger acceptance and efficiency shown in Fig.~\ref{fig:hadron_effic}, the total usable hadronic rate becomes 14.2 kHz.

\subsection{Data analysis}\label{sec:analysis}

\begin{figure}[ht!]
\center 
\includegraphics[width=0.55\textwidth]{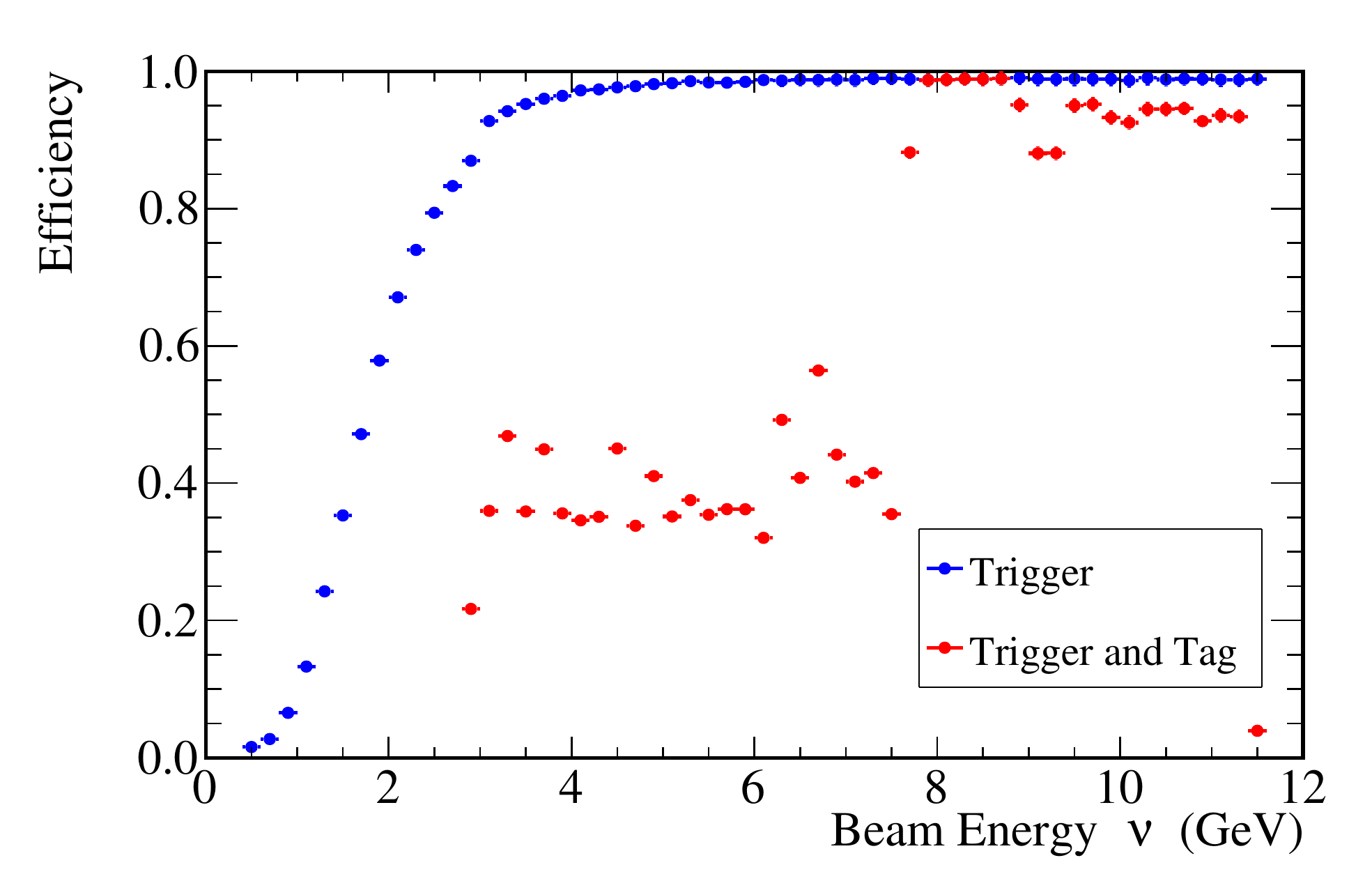}
\caption{\small{\label{fig:hadron_effic} Simulation of hadronic events in the GlueX apparatus. The blue symbols show the average efficiency for triggering on these events with the standard GlueX trigger, plotted versus the photon energy. The efficiency accounting for tagging the energy of the photon with the tagger is shown by the red symbols. See Sec.~\ref{sec:simulation} for details.}}
\end{figure}

The data analysis will be done inclusively by counting triggered events matched with a tagged beam photon.  
Although the detector does not have perfect acceptance and efficiency to detect individual particles, when analyzed in an inclusive fashion the combination of acceptance and efficiency for an \emph{event} approaches unity.  Simulation shows that the trigger efficiency for hadronic interactions is $\varepsilon>93\%$ for $\nu>3$\,GeV and $\varepsilon>98\%$ for $\nu>5$\,GeV, requiring only a small correction, see Fig.~\ref{fig:hadron_effic}.  The efficiency for tagging is primarily geometrical and is measured to the 1\% level in dedicated runs.  See Sec.~\ref{sec:simulation} for details on the simulations.

\begin{figure}[ht]
\center 
\includegraphics[width=0.6\textwidth]{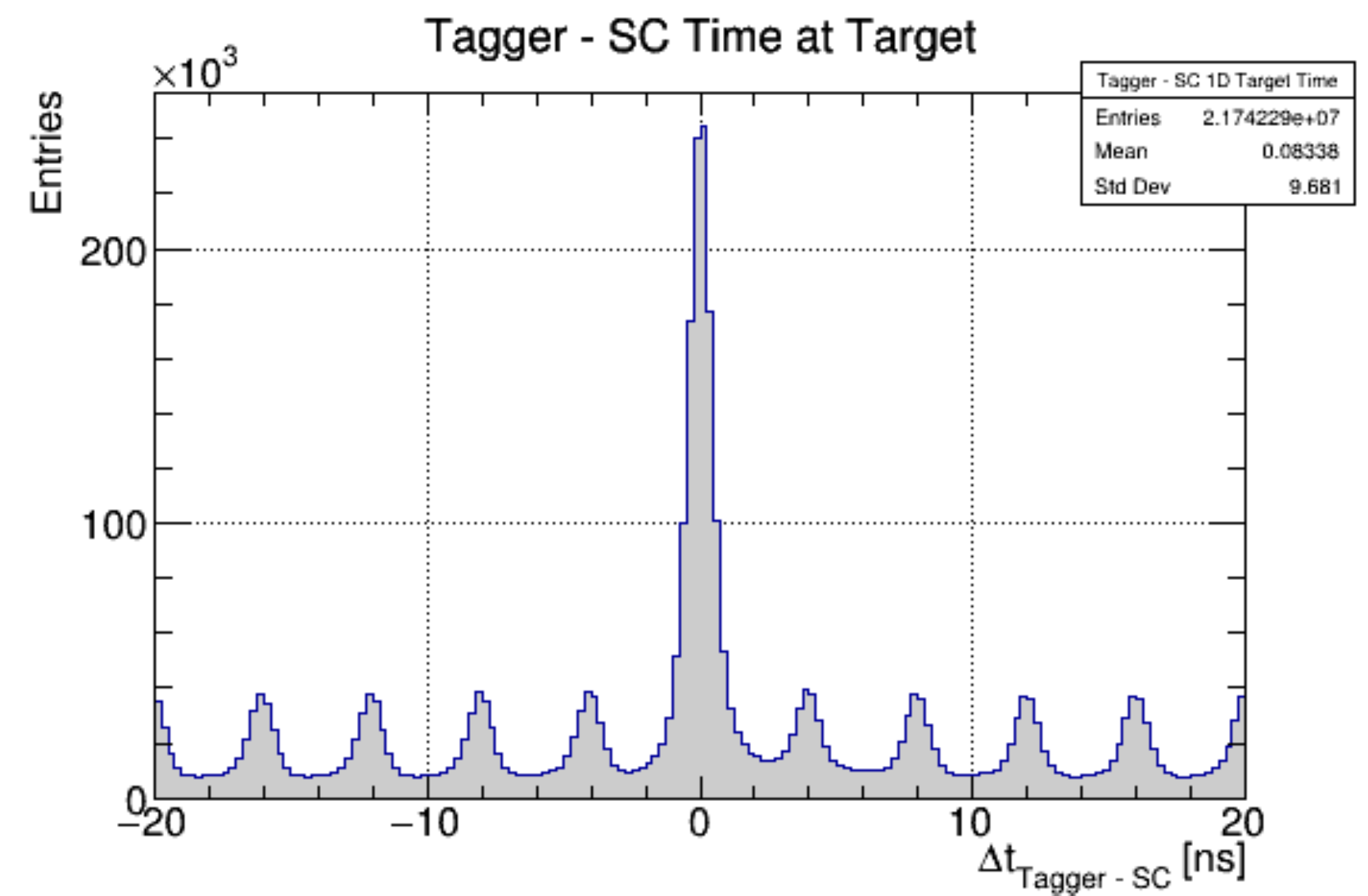}
\caption{\small{\label{fig:tagger_accid2} Experimental data for the timing difference between the Hall D start counter and tagger,
with a 300 nA beam on a 1.86$\times 10^{-5} X_0$  aluminum radiator.  The prompt peak compared to the out-of-time peaks 
indicates that the accidentals under the prompt peak contribute just less than 15\%.}}
\end{figure}

For a triggered event, the energy of the event will be determined by correlating with an in-time photon in the tagger using coincidence timing.  At higher luminosity it is possible to have more than one signal in the tagger for a given beam bunch.  
This accidental background will be subtracted statistically using known techniques, leading to a statistical penalty due to the measurement of the background.
The expected running conditions, 240 nA on a 1.86$\times 10^{-5} X_0$  radiator,  produce an  accidental background less than 15\%, see Fig.~\ref{fig:tagger_accid2}.

Assuming that a fraction $f$ of the events in the prompt (in-time) peak are accidental, we can measure this contribution using the side (out-of-time) peaks adjacent to the prompt peak.  The resultant statistical uncertainty is given by
\begin{equation}
    \sigma = \sqrt{N_p+\frac{fN_p}{M}} = \sqrt{N_p}\sqrt{1+\frac{f}{M}},
\end{equation}

\noindent where $N_p$ are the prompt events, $fN_p$ the number of accidental events in any peak and $M$ the number of side peaks used in the subtraction.  For $f=0.15$, and using $M=10$ side peaks, the fractional increase in the statistical uncertainty due to the subtraction is 0.7\%.

\subsubsection{Convergence of the integral}

Analysis of the integral convergence  requires measuring the $\nu$-dependence of $\Delta\sigma(\nu)$, i.e. $N^+ - N^-$, but not necessarily the absolute normalization.  This is achieved most easily by simply counting the number of triggers.  Unpolarized backgrounds will cancel in the difference. Any significant polarized background (from Bethe-Heitler pair production) can be corrected for.

\subsubsection{Validity of the sum rule}

In contrast to checking the $\nu$-dependence of $\Delta \sigma(\nu)$, performing a measurement of the GDH integral requires an absolute polarized cross-section measurement, which will be obtained by normalizing the yield by the beam flux, target density, solid angle, target and beam polarizations, and efficiencies.

As verification of the primary analysis method, an asymmetry analysis strategy is also possible.  The relative asymmetry  $A=(N^+ - N^-)/(N^+ + N^-) \equiv {\Delta \sigma(\nu)}/{2\sigma(\nu)}$ can be formed and, together with the well-measured $\sigma(\nu)$ shown in Fig.~\ref{fig:totalCS}, one can obtain the absolute $\Delta \sigma(\nu)$. 

For an analysis of the relative asymmetry  --and in contrast to the $\Delta \sigma$ analysis-- dilution by unpolarized target material must be corrected for.  To provide the possibility of such analysis, a $^{12}$C foil will be placed near the FROST cell, see Section~\ref{target}.  It will allow to estimate the dilution.  Empty target runs will also be necessary.

\subsubsection{Exclusive analysis}\label{sec:exclusiveanalysis}

In addition to the inclusive analysis, a fully exclusive analysis will also be done to study the background from electromagnetic processes and to verify that the acceptance and efficiency are well understood.  

Only a fraction of the events can be studied exclusively since it requires detecting all the final state particles.  In an exclusive analysis, the measured energies of the final state particles provide an independent measure of the beam photon energy, allowing a careful check of the primary analysis method.

We anticipate studying, at a minimum, the background processes of Bethe-Heitler pair production and Compton scattering.  As examples of hadronic reactions, the detection and trigger efficiency for minimally ionizing particles can be studied with $\gamma p\to p\rho$; neutrons with $\gamma p\to n\pi^+$ and photons with $\gamma p\to p\pi^0$.  As a by-product of this analysis, the spin-dependent cross-sections of a number of different exclusive reactions are likely to be produced for the first time at high-energy.  The overall analysis will benefit greatly from the extensive work done by the GlueX Collaboration to understand the efficiency and acceptance of the detector in the time leading up to the experiment.

\subsection{Backgrounds}
\label{sec:backgrounds}

Backgrounds to the signals of interest to this proposal ($\Delta y (\nu)$ and  $\Delta \sigma (\nu))$ are those that do not cancel when taking the helicity difference, \emph{viz} the spin-dependent or ``polarized" backgrounds.  Unpolarized backgrounds cancel in the yield or cross-section difference. They can impair the experiment only by contributing to the total DAQ rate: If the DAQ rate limit (presently 80 kHz) is reached due to the extra rate from unpolarized backgrounds, it will reduce the statistical precision of the experiment. This is indeed the case but as we will show, the statistics remain more that sufficient.

\subsubsection{Polarized backgrounds from the FROST target}

Two possible sources of polarized backgrounds exist: electromagnetic 
(Compton scattering, $ \overrightarrow\gamma \overrightarrow{e} \to \gamma e  $, 
and Bethe-Heitler process, $ \overrightarrow\gamma \overrightarrow{p}\to e^+ e^-p$) 
and hadronic (polarized scattering from non-hydrogen, or non-deuterium, 
nuclei) backgrounds.  However, the spin-dependent part of these backgrounds is not expected 
to be significant with the FROST target, as explained below.

\paragraph{Compton scattering}

There should be no significant polarized Compton
background because all the electrons of the $^{16}$O and $^{12}$C 
nuclei making the butanol of the target are spin-paired. In addition, 
the single (unpaired) electrons of the hydrogen atoms  will also 
be unpolarized since each is shared with the $^{16}$O of the OH-bond. 
There will be a small polarization of the electrons of $^{16}$O 
and $^{12}$C if some of them become ionized. The remaining unpaired
electrons of the ionized nuclei will be fully polarized in the direction 
of the solenoid field. However, the fraction of such electrons is estimated 
to be at less than the $10^{-4}$ level \cite{CKeith_PC}.
 
Furthermore any polarized Compton contribution to the GDH sum, $\int \Delta \sigma\,\mathrm{d}\nu$, will be nearly fully suppressed (and exactly suppressed for an experiment with perfect detector efficiencies and solid angle coverage) since the GDH sum rule on the electron is expected to be zero and to have fully converged to this null value by $\nu=0.5$ GeV~\cite{Pantforder:1998nb}, well below our lower energy acceptance of 3 GeV. We remark that in the case of the electron, the sum rule prediction is much more solid than for the nucleon because it only involves QED which is well-known at the energies of this proposal and far beyond.

Finally, flipping the target spin direction will cancel this already 
 completely  negligible background.

\paragraph{Bethe-Heitler process}

\begin{figure}[hbtp]
\begin{center}
\includegraphics[width=15cm]{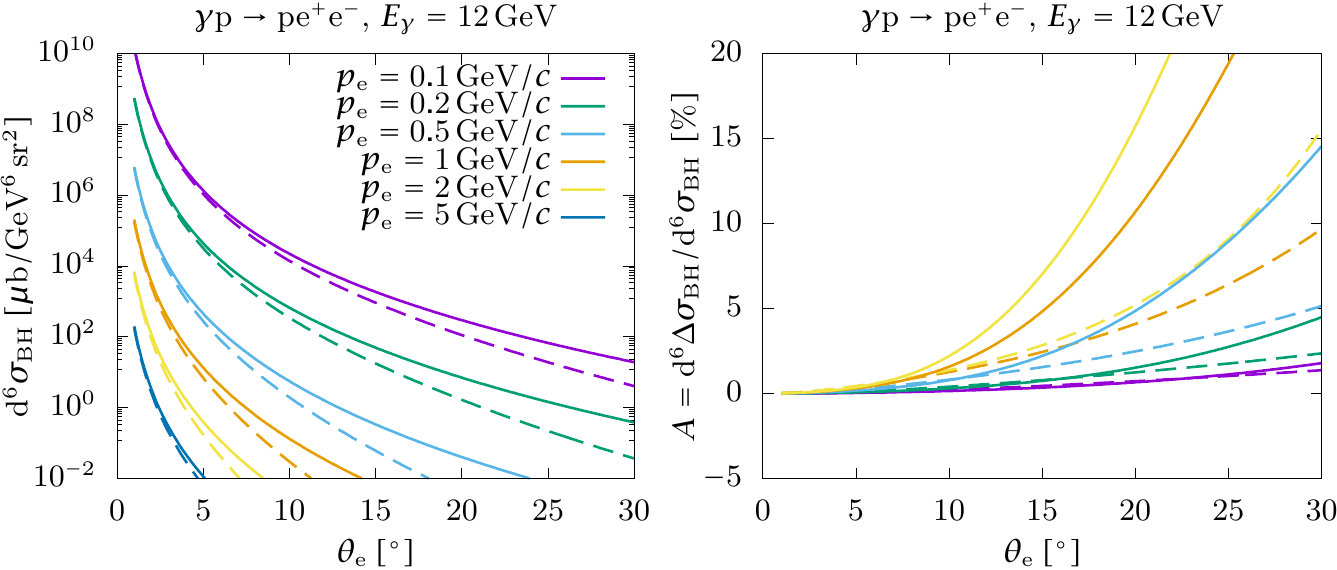}
\end{center}
\vspace*{-5mm}
\caption{{\small Left: elastic (dashed curves) and elastic+inelastic (full curves) Bethe-Heitler ($e^+e^-$) unpolarized cross-section
for $E_\gamma = 12\,\mathrm{GeV}$ as a function of the scattering angle.  Right: Bethe-Heitler photon-target asymmetry (same curve notation).  Note that while the BH asymmetries
appear to be substantial, they need to be multiplied
by the unpolarized cross-section to yield the BH contribution  
$\Delta\sigma(\nu)$, yielding with the standard GlueX trigger configuration a (negligible) few tens of $\mathrm{n}\mathrm{b}$ at most: see Fig.~\ref{fig:BH_spinCS}.}}
\label{fig:BHe}
\end{figure}

The Bethe-Heitler (BH) background can 
be spin-dependent. Only the pair-created leptons scattering off 
the polarized protons or deuterons produce an asymmetry, 
since the $^{16}$O and $^{12}$C nuclei are unpolarized.
We are thus not concerned here with BH on those nuclei.  
For a purely nucleonic target the unpolarized and polarized 
BH cross-section and the corresponding asymmetry can be evaluated 
explicitly: see \cite{Tsai74,Gehrmann:1997qh} and references therein.  
As an illustration, Fig.~\ref{fig:BHe} shows the calculated six-fold 
differential cross-section for $e^+e^-$ production on the proton
with $E_\gamma = 12\,\mathrm{GeV}$, in the simplified case 
$p_{e^-} = p_{e^+} \equiv p_e$, as well as the photon-target asymmetry, 
as a function of the scattering angle $\theta_{e^-} = \theta_{e^+} \equiv \theta_e$.
In the calculation of the inelastic contributions 
to the polarized BH cross-sections which were used 
to compute the corresponding asymmetries,
we have used the recent and stable parameterizations 
of $g_1^p(x,Q^2)$ and $g_2^p(x,Q^2)$ spin structure 
functions published in Ref.~\cite{Simula2002}.

The $e^+e^-$ BH process is expected to dominate over the analogous 
$\mu^+\mu^-$ BH process.
If the BH process leaves 
the nucleus intact, the background can be corrected nearly exactly 
since the well-known nucleon electromagnetic form factors are 
the only phenomenological inputs necessary.  

\paragraph{Hadronic background} 

There will be no double-spin difference or asymmetry arising from polarized hadronic 
background because all nucleons in $^{16}$O and $^{12}$C nuclei are spin-paired.

\subsubsection{Spin-independent backgrounds}
\label{spinindepBG}

Although unpolarized backgrounds cancel exactly in $\Delta \sigma$, they still affect the experiment as they can consume some of the 80 kHz DAQ capability. 
The hadronic background rate 
is estimated to be 31 kHz, assuming a 120 $\mu$b $\gamma N$ cross-section, a 10 cm target length, a 3 mm thick carbon foil and the tagger efficiency shown in  Fig.~\ref{fig:hadron_effic}.

Spin-independent backgrounds arise from photoproduction, from Compton scattering, and from the Bethe-Heitler process on $^{12}$C and $^{16}$O.  These have been simulated with {\footnotesize{GEANT}}, see Sec.~\ref{sec:simulation} for details.  Compton scattering  was simulated for hydrogen and then scaled by the number of protons in the target.  The Bethe-Heitler process was simulated for the nucleon and scaled up using the target dilution factor. 
These rates will be confirmed by
analyzing GlueX data on the Kapton ([C$_{22}$H$_{10}$N$_2$O$_5$]$_n$) 
windows of the unpolarized LH$_2$ target.
In all, the 80 kHz data acquisition limit is shared between a total hadronic rate of 35.9 kHz, a 35.8 kHz BH rate and a 8.3 kHz Compton rate, see Fig.~\ref{Fig:XS and rates}. This reduces the number of hadronic events by a factor of 2 compared to the case where there would be no background. 
The situation can be improved by upgrading the DAQ and refining  the trigger. Although desirable, this is not necessary since the present hadronic rate is already enough to provide sufficient statistics
within a week (10 days) of running on the proton (deuteron).

\section{Low-energy running}

As will be shown in Section~\ref{sec:uncert_stats}, if the experiment is performed only at the nominal CEBAF beam energy, there will remain a substantial gap in $\nu$ coverage, between 1.8\,GeV and $3$\,GeV, for the deuteron and thus the neutron---and therefore also for the very desirable isospin decomposition.

We therefore propose to take data with a beam energy between \sfrac{1}{2} and \sfrac{1}{3} of the nominal beam energy. This would bring the lower reach of the experiment down to 1.0--1.5\,GeV.  Any energy in that range would be suitable. As an example, we will assume an electron beam energy 4 GeV in this document.
The benefits of such a data set would be extensive:
\begin{itemize}
    \item we would smoothly link the existing neutron world data with our Hall D data, with no energy gap;
    \item we would obtain an overlap between our measurements at two different beam energies, in the 3\,GeV to 4--6\,GeV region using different regions of the tagger and different absolute polarizations of the photon beam;
    \item it would allow us to study the fourth and even third resonance regions for both the proton and the neutron;
    \item there would be an overlap with existing proton and neutron data to significantly improve on the statistical and systematic uncertainties on the GDH sum;
    \item we could improve the determination of the Regge parameters even further and determine the minimum energy at which the Regge phenomenology becomes applicable. 
\end{itemize}

While it would facilitate the analysis to take the low energy data within the same run period as that of the nominal energy data, this might be difficult to schedule. A separated low energy run, e.g. during summer as it is done frequently, would achieve the same goal since the low and high energy runs have a partial energy overlap, thereby offering   a means to normalize out any change in the detector performance that would shift $\Delta y(\nu)$.

\section{Geant simulation of the experiment \label{sec:simulation}}

The experiment has been simulated using the same detailed and well tested simulation chain used in GlueX.  Event generators specify one or more primary vertices to be simulated, which are randomized within the target with timing that matches the RF structure of the beam.  A simulation code, either {\em hdgeant}\footnote{\url{https://github.com/JeffersonLab/halld_sim}} or {\em hdgeant4}\footnote{\url{https://github.com/JeffersonLab/HDGeant4}}, tracks the particles through the experimental setup and records the signals they produce in the active elements of the detector.  The output of the simulation is further processed to account for detector inefficiencies and resolutions and to overlay additional hits from uncorrelated background events.  The simulation 
uses the same geometry definitions and magnetic field maps as used in real events reconstruction.  The geometry includes the full 
photon beamline, from the radiator to the photon dump.
The simulated events are then processed with the same reconstruction software as used for the real events~\cite{Adhikari:2020cvz}.

Simulations have been done of the trigger acceptance and tagging efficiency  for various processes, considering whether the standard GlueX trigger would fire for each and whether the electron which radiated the beam photon would be tagged.  The tagging efficiency below about 7.8\,GeV drops to about 45\% because the tagger is designed only to sample this region of the spectrum rather than detect all electrons.

\begin{figure}[htb]
\center 
\includegraphics[width=0.75\textwidth]{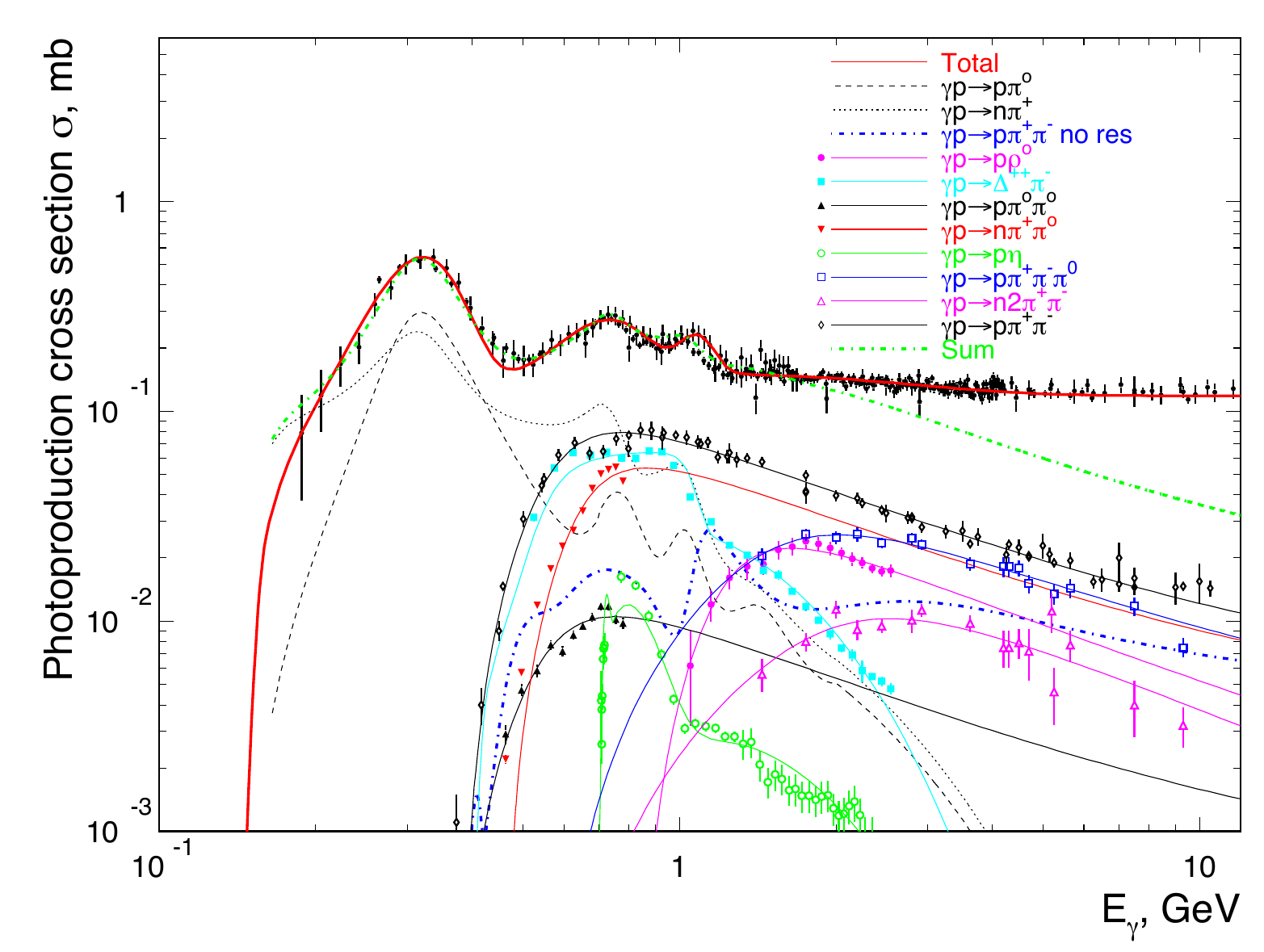}
\vspace{-0.4cm}
\caption{\label{fig:bggenmodel}\small{For energies below 3\,GeV, a model, comprising 11 processes with low multiplicity, is able to describe both the total cross-section and the individual cross-sections.}}
\end{figure}

Fig.~\ref{fig:hadron_effic} shows the acceptance for the hadronic events that are the signal in this experiment.  The hadronic event generator, called {\em bggen}, is a standard and well tested tool in the GlueX analysis.   It is based on Pythia~\cite{Sjostrand:2006za},
but includes additions that describe the low-energy photoproduction cross-sections in the resonance region.  Fig.~\ref{fig:bggenmodel} shows the model used for $\nu<3$\,GeV.  The simulation demonstrates a high trigger efficiency for these events.  The average trigger and tagging acceptance over the 3\,GeV to 12.0\,GeV spectrum is 55\%.  

Bethe-Heitler events have a large total cross-section in the range 12 to 15 millibarns at these energies.  The majority of these events are produced with leptons at very small lab angles and the GlueX trigger effectively suppresses this rate to a manageable amount.  

Simulations of the Bethe-Heitler events were done using the \emph{Diracxx} software package.\footnote{\url{https://github.com/rjones30/Diracxx}}  This is a general-purpose toolkit for use within the CERN/ROOT framework for computing cross-sections and rates with all polarization observables under the control of the user for both incoming and outgoing particles.  In the simulation, the Bethe-Heitler process is treated in a fashion fully consistent with tree-level QED, taking into account the polarization of the photon and both space-like and time-like form factors of the proton target. Internal radiative corrections are not currently included, but external radiation is automatically taken into account by the Geant4 tracking library.  

\begin{figure}[ht]
\center 
\includegraphics[width=0.55\textwidth]{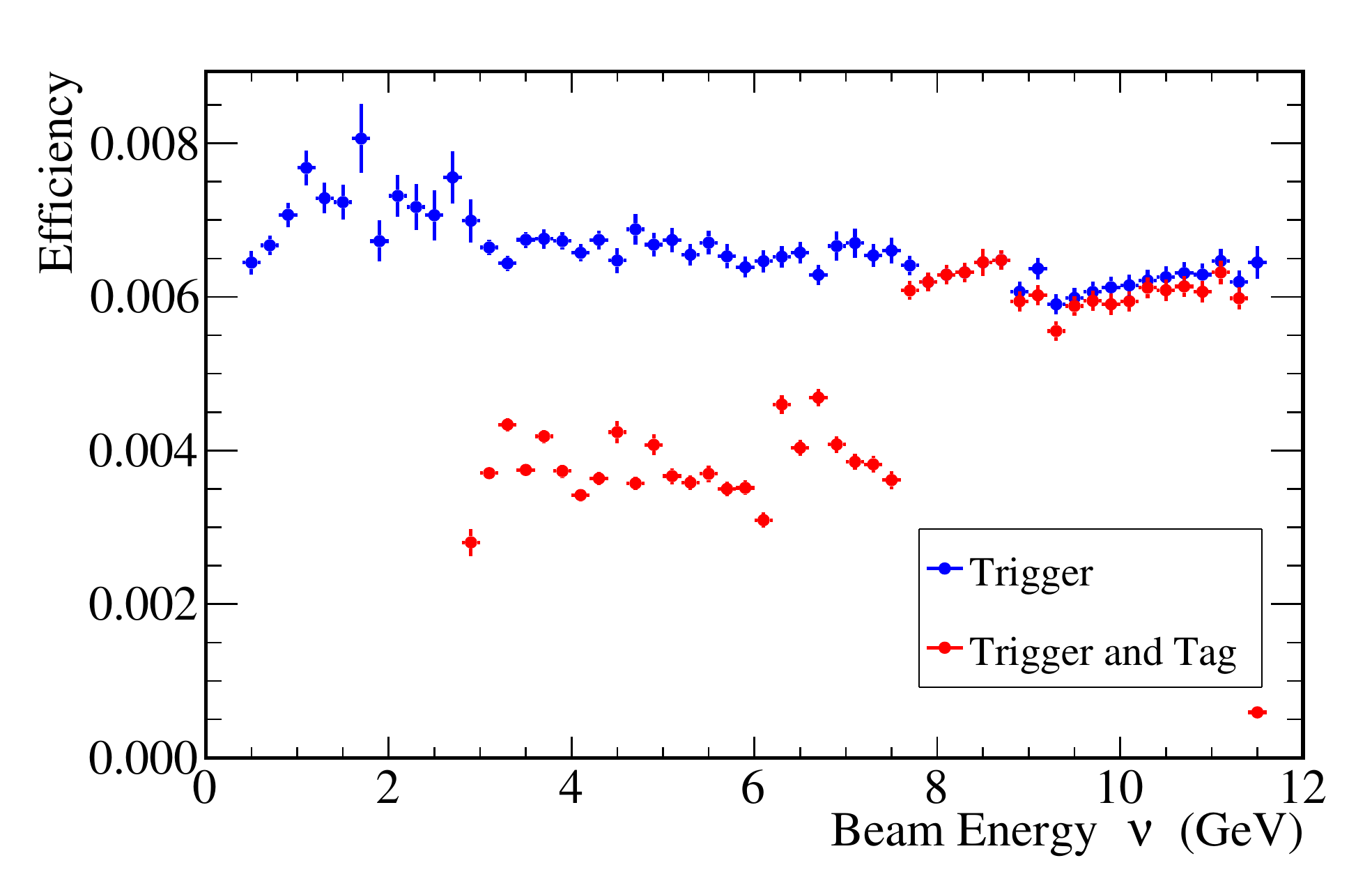}
\caption{\small{\label{fig:BH_effic} Average efficiency for triggering on Bethe-Heitler $e^+e^-$ events as a function of photon beam energy accounting for tagging the energy of the photon with the tagger (red symbols) and without accounting for the tagger efficiency (blue symbols).}}
\end{figure}

Fig.~\ref{fig:BH_effic} shows the efficiency for triggering on and tagging Bethe-Heitler events using the standard GlueX trigger.  Use of the trigger brings the rate down to the same order as the hadronic triggered rate.

\begin{figure}[ht]
\center 
\includegraphics[width=0.55\textwidth]{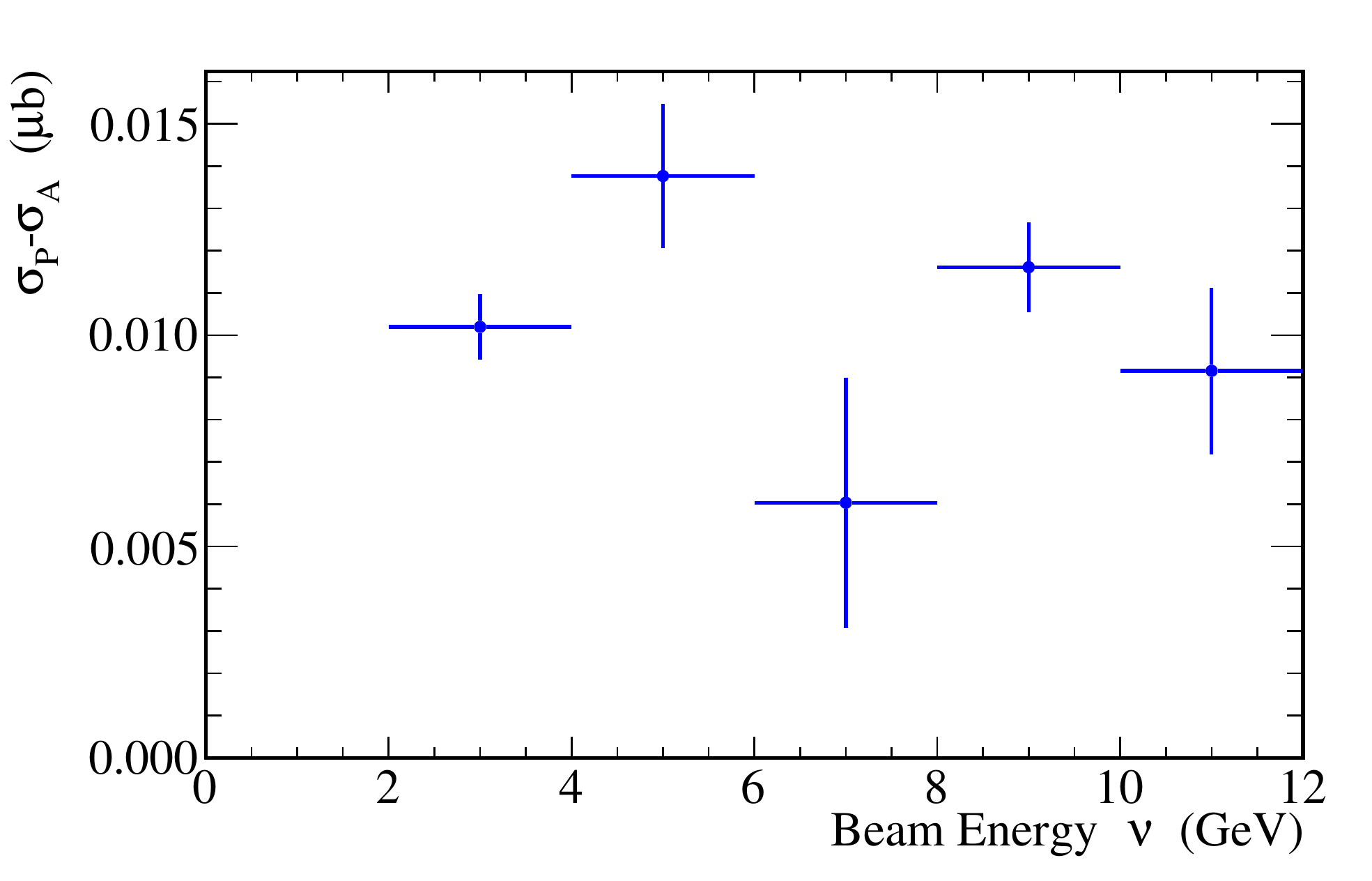}
\caption{\small{\label{fig:BH_spinCS} Spin-dependent total cross-section for Bethe-Heitler $e^+e^-$ events which fire the standard GlueX trigger, plotted as a function of photon beam energy.}}
\end{figure}

Fig.~\ref{fig:BH_spinCS} shows the polarized total cross-section for Bethe-Heitler $e^+e^-$ events which fire the standard GlueX trigger and have the beam photon tagged (red events in Fig.~\ref{fig:BH_effic}).  On average these events have a spin-dependent total cross-section less than 20~nb which less than 1\% of that expected  for the hadronic events (Figs.~\ref{fig:expect}, \ref{fig:expect_n} and \ref{fig:expect_D}).

\begin{figure}[ht]
\center 
\includegraphics[width=0.55\textwidth]{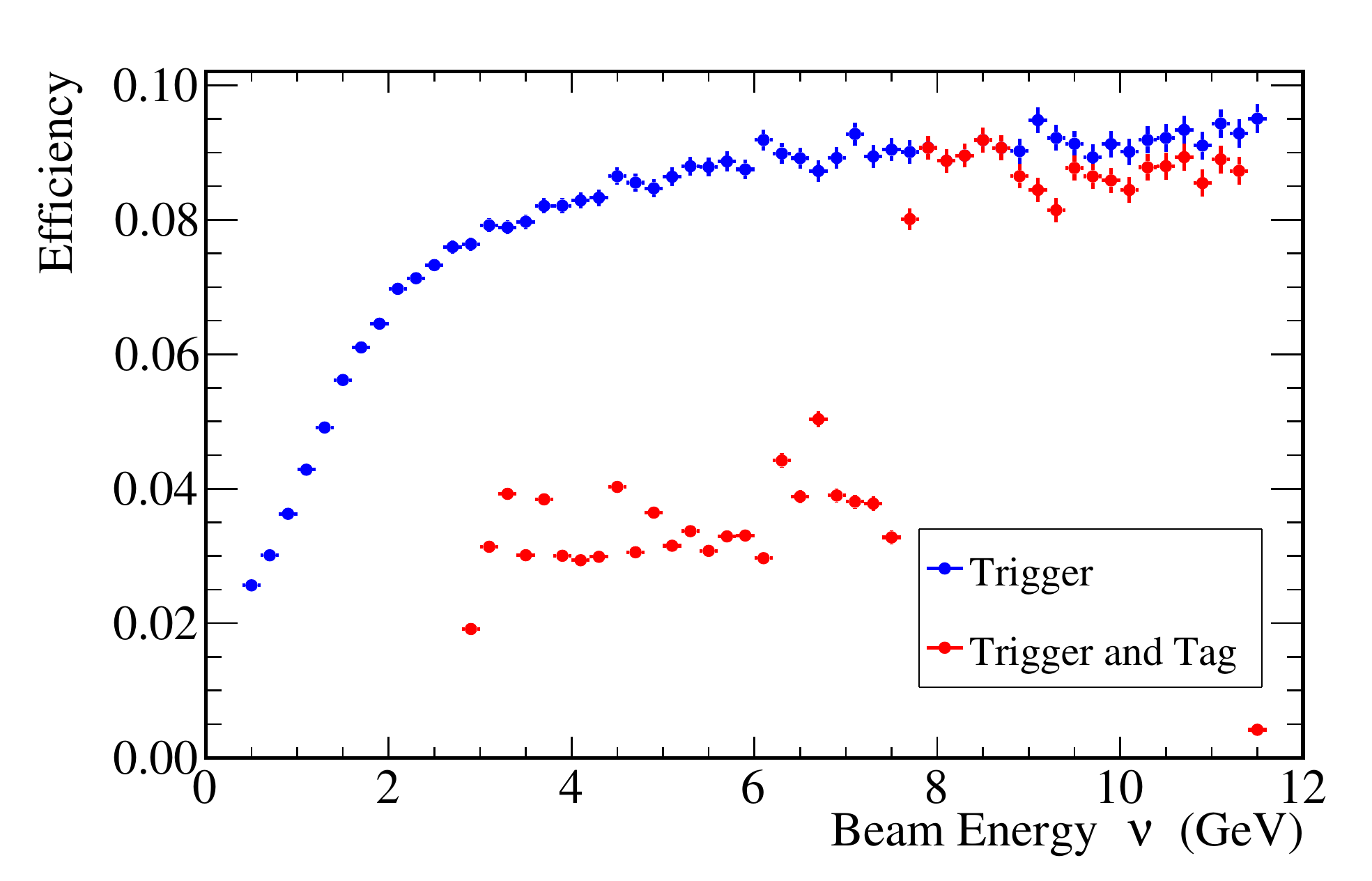}
\caption{\small{\label{fig:Compton_effic} Average efficiency for triggering on (and tagging) Compton events with the standard GlueX trigger, plotted as a function of photon beam energy.}}
\end{figure}

Fig.~\ref{fig:Compton_effic} shows the acceptance for Compton scattering from atomic electrons, $\gamma e\rightarrow\gamma e$.  Only about 7\% of Compton events trigger the detector and are tagged.

These simulations will further be used to study the relative fractions of the various processes in order to optimize the trigger condition for the experiment to reject background while maintaining high acceptance for the hadronic events of interest.

~

To conclude, the simulation shows that the polarized background contribution is very small and thus, once corrected with the same tools as used here, will be entirely negligible.

\section{Statistical precision and sensitivity of the determination of the $\nu$-dependence of $\Delta \sigma$} \label{sec:uncert_stats}

\subsection{Statistical precision}

To estimate the beam time necessary for the measurement, we use   a total collimated photon flux of $7 \times10^7$ s$^{-1}$. Such flux can be obtained with the currently available 
$1.86\times10^{-5} X_0$  aluminum radiator (1.64\,$\mu$m), 240\,nA electron beam current and the standard 5\,mm collimator.  To determines the trigger rate, we use:
 
$\bullet$ the flux between $\nu=3$ and 12\,GeV: $1\times10^7$ s$^{-1}$ for the nominal energy run and $2.5\times10^6$ s$^{-1}$ between $\nu=1$ and 4\,GeV for the low energy run\footnote{for the low energy run, the beam size contraction due to the Lorentz boost effect is smaller. The consequent lower photon transmission through the Hall D main collimator is accounted for by a reduction of 10/28 in the photon flux compared to the nominal case.}.

$\bullet$ a 80\% for the detector/trigger efficiencies above $\nu=7.5$~GeV ($\nu=2.5$~GeV)
and 40\% below (see Fig.~\ref{fig:BH_effic}) for the nominal (low) CEBAF energy run.

$\bullet$  a 80\% for the electron beam and target polarizations. \\
For $\Delta \sigma$, we use the Regge form 
\begin{equation}
\sigma_P-\sigma_A=Ic_1 s^{\alpha_{a_1}-1} +c_2 s^{\alpha_{f_1}-1},
\label{eq:Regge_expect}
\end{equation}
with $s=2M\nu + M^2$, $I=\pm$ is the isospin sign of the proton or neutron, and with values $c_1=-34.1~\mu b$, $\alpha_{a_1}=0.42$, $c_2=209.4~\mu b$, $\alpha_{f_1}=-0.66$~\cite{Helbing:2006zp}.

\subsubsection{The 12 GeV run}

\begin{figure}
\center 
\includegraphics[width=0.45\textwidth]{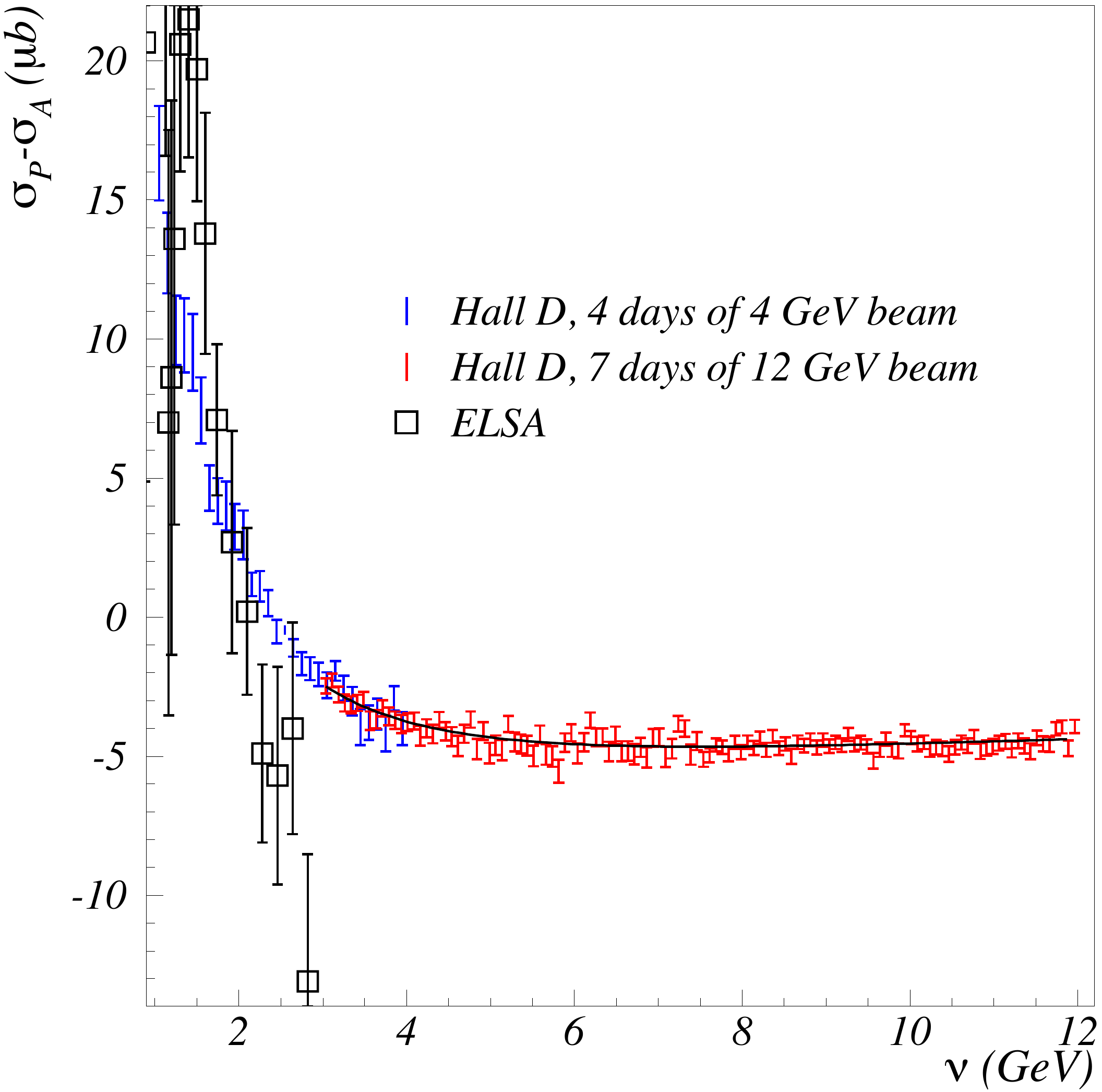}
\includegraphics[width=0.45\textwidth]{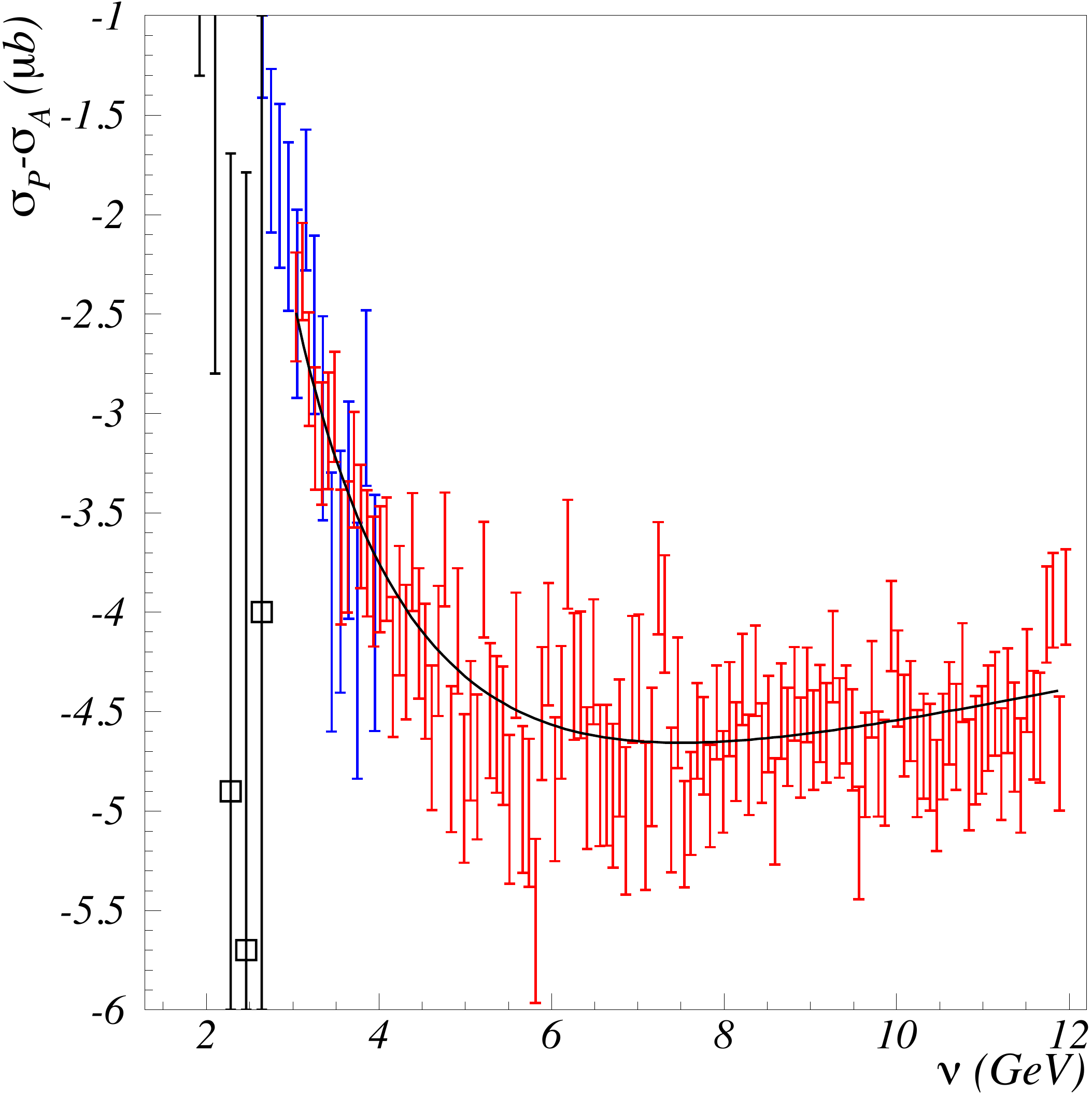}
\caption{\small{\label{fig:expect}\small{Left: $\Delta \sigma$ on the proton from ELSA high-$\nu$ data (squares) and expected results from Hall D using a 12 GeV beam (red) and a 4 GeV beam (blue). The plain line is the best fit to the  simulated 12 GeV data shown in red and based on the Regge form of Eq.~(\protect\ref{eq:Regge_expect}). It yields 
$\alpha_{a_1}=0.418 \pm0.009$ and $\alpha_{f_1}=-0.614\pm0.037$ 
for the intercepts of the $a_1$ and $f_1$ Regge trajectories. 
Only the statistical uncertainty is relevant to determining the intercept values. The systematic uncertainties, expected to be at the 5\% level, are not shown.
Right: zoom on the expected Hall D data. 
}}}
\end{figure}

\begin{figure}
\center 
\includegraphics[width=0.45\textwidth]{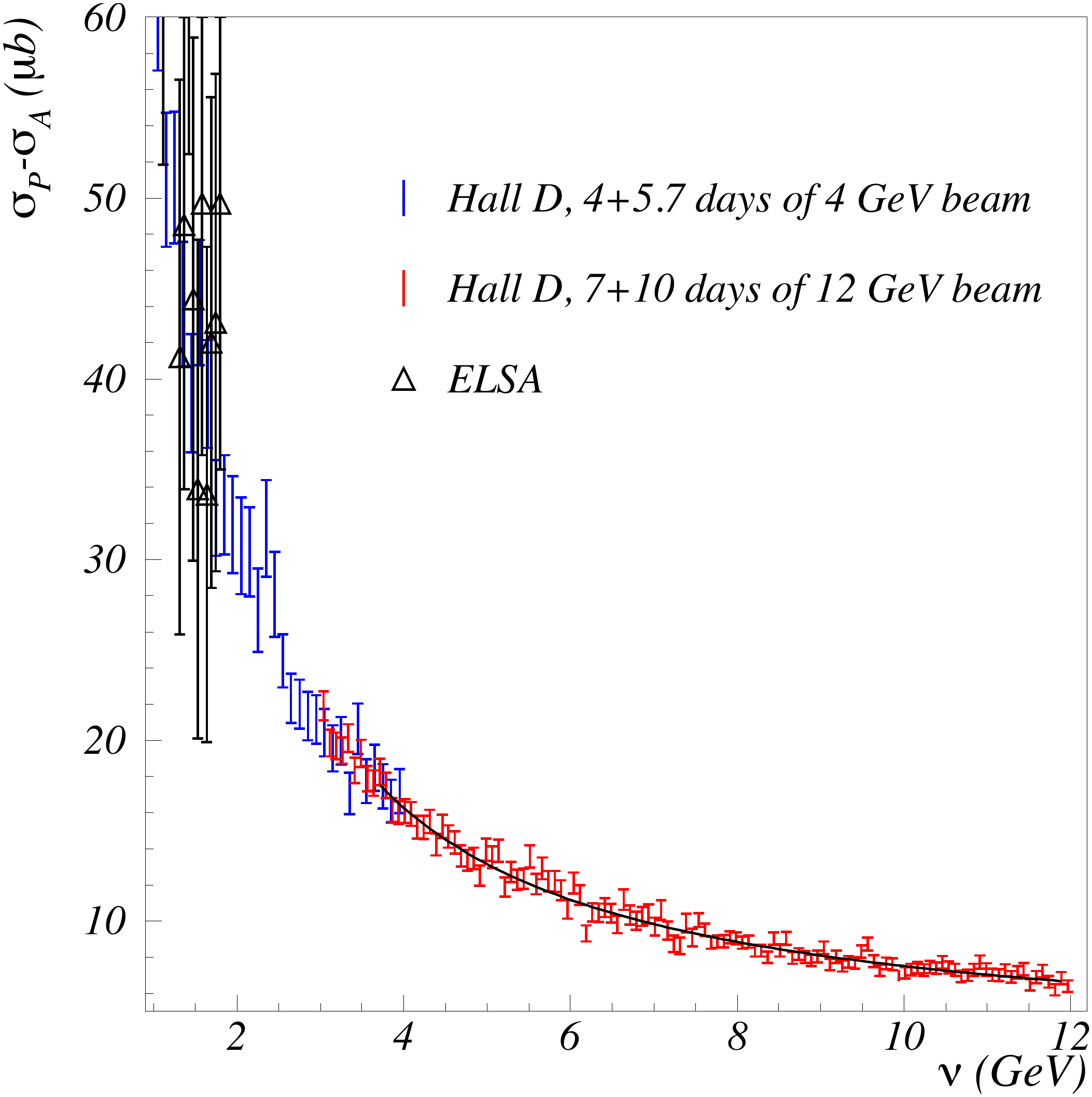}
\includegraphics[width=0.45\textwidth]{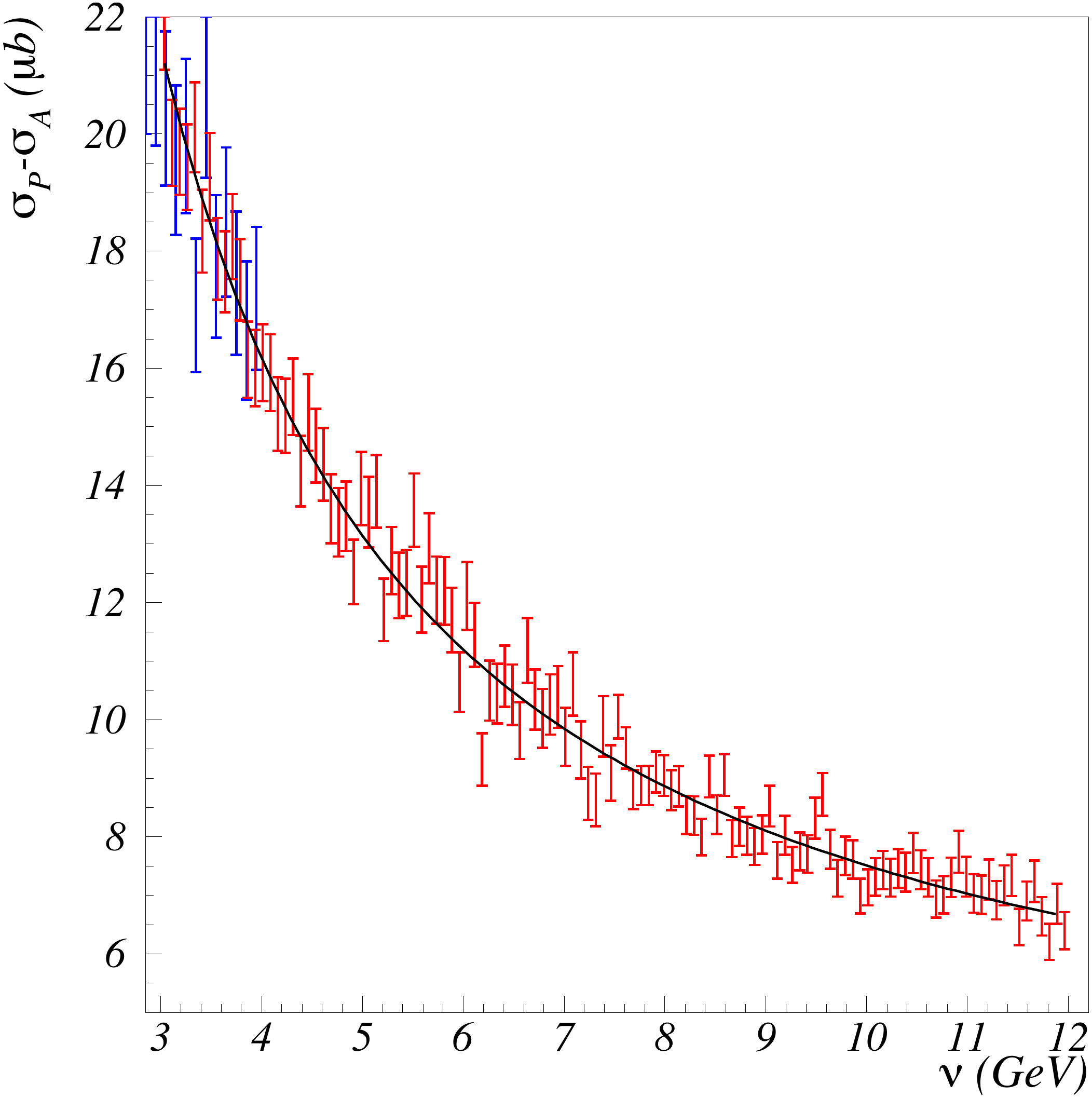}
\caption{\small{\label{fig:expect_n}\small{Same notation
as in Fig.~\protect\ref{fig:expect}, but for the neutron extracted from deuteron data (statistical uncertainty only, the systematic ones being unimportant). 
The best values for the intercepts of 
the $a_1$ and $f_1$ Regge trajectories are 
$\alpha_{a_1}=0.412 \pm0.013$ and $\alpha_{f_1}=-0.629\pm0.062$.
}}}
\end{figure}

\begin{figure}
\center 
\includegraphics[scale=0.4]{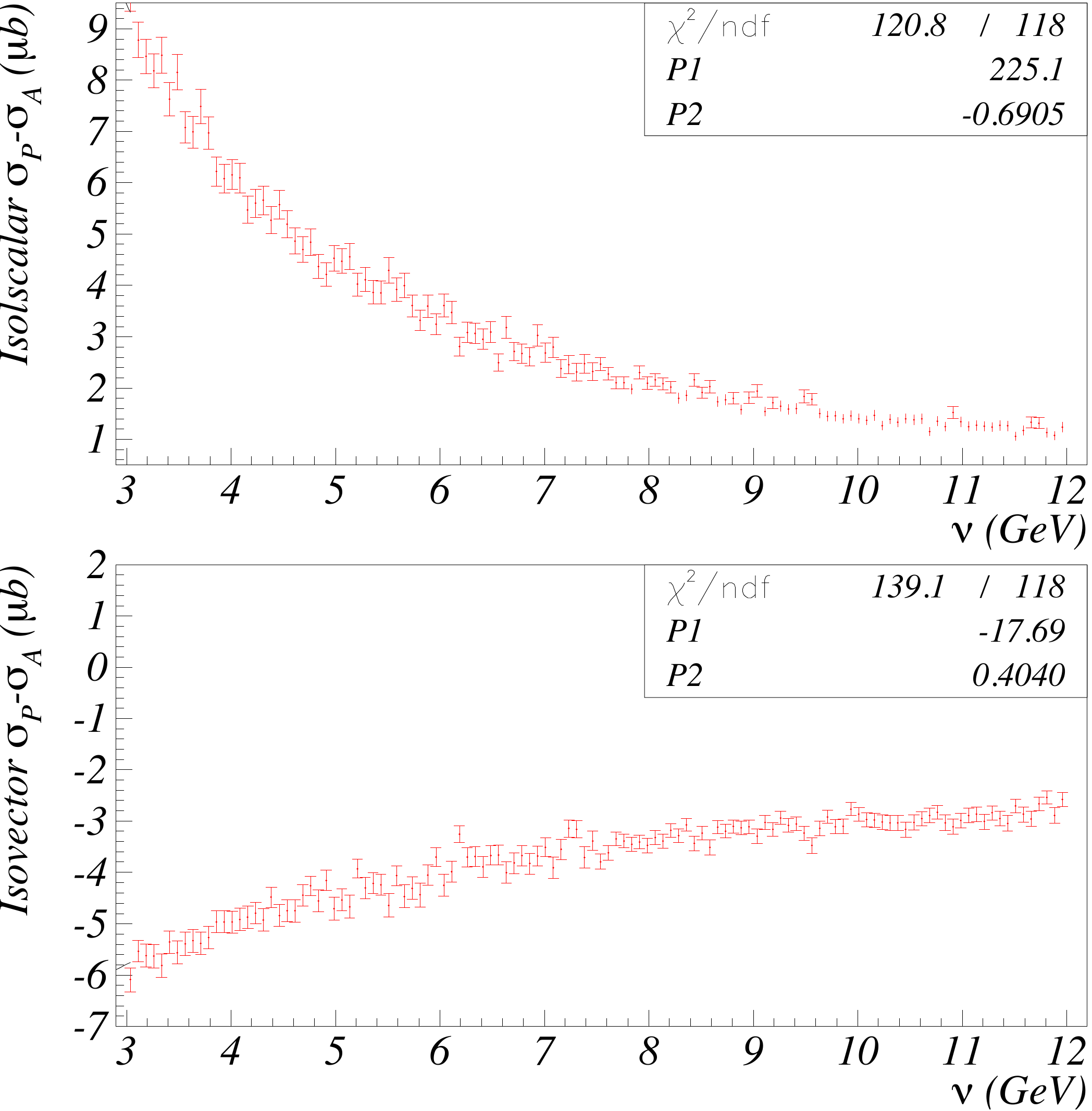}
\caption{\small{\label{fig:isospin}\small{Isospin decomposition of  $\Delta \sigma$. 
Top: isoscalar part, with best value for the intercepts of the $f_1$-meson Regge trajectory $\alpha_{f_1}=-0.691 \pm0.029$}. 
Bottom: isovector part, with best value for the intercepts of the $a_1$-meson Regge trajectory $\alpha_{a_1}=0.404 \pm0.011$.
(Statistical uncertainty only.)}}
\end{figure}

\begin{figure}
\center 
\includegraphics[width=0.45\textwidth]{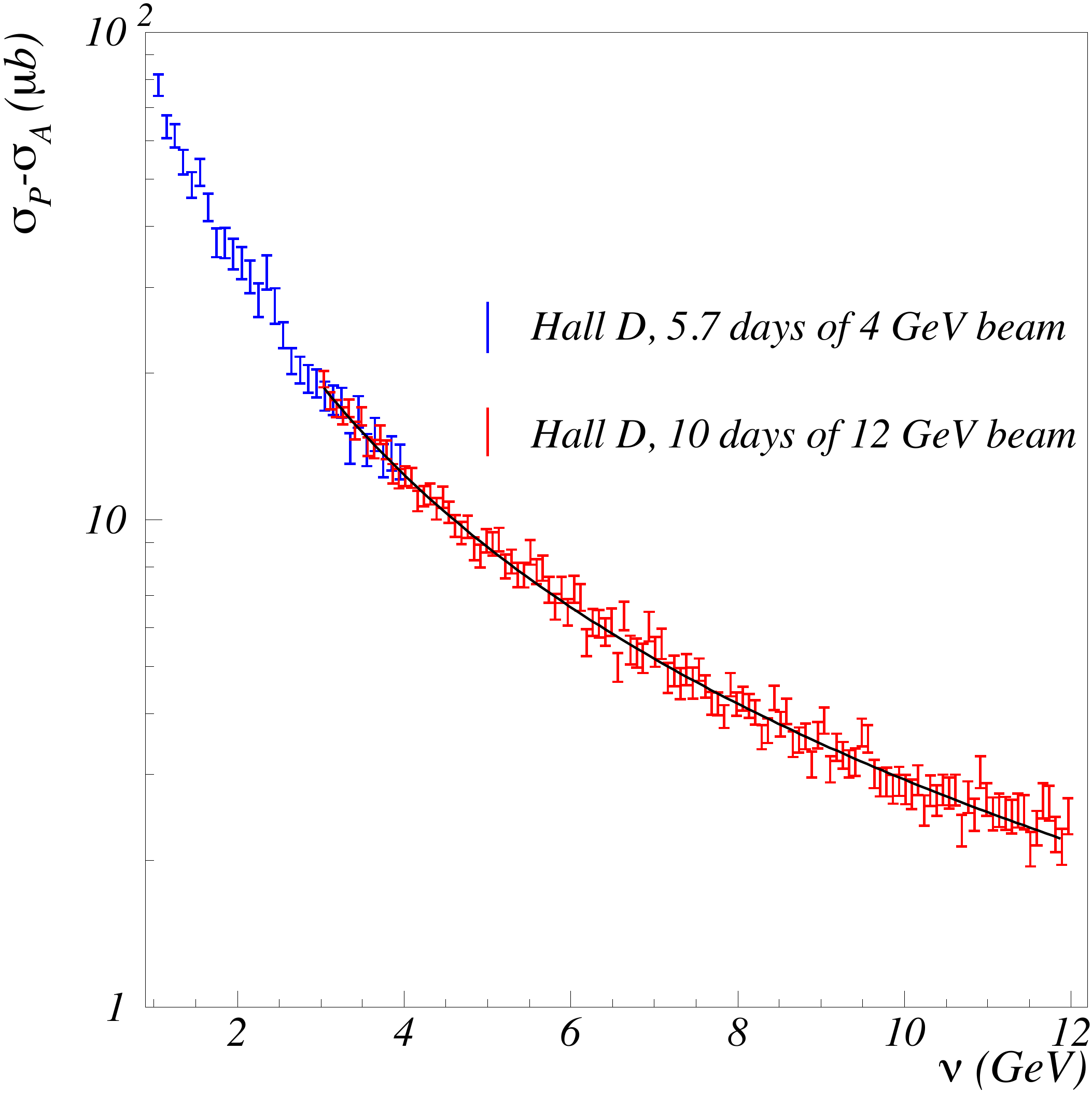}
\caption{\small{\label{fig:expect_D}\small{$\Delta \sigma$ for the deuteron expected results from Hall D (statistics only.). 
The best fit to these simulated data is $\Delta \sigma = 450(34)s^{-0.691(29)}$.}}}
\end{figure}

The highest available CEBAF beam energy is optimal to study the GDH sum rule. We assume 12 GeV will be available and we suppose that one week of running on hydrogen is a minimum given the investment of two months to install the target.
If $\Delta \sigma$ indeed follows Eq.~(\ref{eq:Regge_expect}), then running 7 days on the proton target and 10 days on the deuteron target 
yields a similar statistical precision for the neutron and proton data, see Figs.~\ref{fig:expect} and \ref{fig:expect_n}, and allows for an optimal isospin analysis, see Fig.~\ref{fig:isospin}. 
The neutron information obtained from the deuteron and proton data can be extracted straightforwardly: at our large $\nu$, in the smooth continuum 
region past the resonances, the deuteron binding (2 MeV) and Fermi motion (115 MeV) can be ignored. The usual formula to extract the neutron information, $n=D/(1-\frac{3}{2}\omega_d)-p$ with $\omega_d=5.6\%$  the probability of the deuteron to be in a $d$-state, is expected to be valid. Hence, no issue regarding nuclear effects is expected for the isospin separation.  The expectation for the deuteron is shown in Fig.~\ref{fig:expect_D}.

This simulation yields statistical%
\footnote{As discussed, correlated systematic
uncertainties causing a global offset of the yields do
no contribute the total uncertainty.}
uncertainties on the intercepts of $\Delta \alpha_{a_1}= \pm0.007$ and $\Delta  \alpha_{f_1}=\pm0.029$. 
These expectations can be compared with the values $\Delta \alpha_{a_1}= \pm0.23$ and $\Delta  \alpha_{f_1}=\pm0.22$ extracted from the ELSA data~\cite{Helbing:2006zp}.
We are comparing here to results from the best fit to the photoproduction data. The intercept values can also be obtained from low-$Q^2$ electroproduction data, and with higher statistical precision, see e.~g.~the recent determination  $\alpha_{a_1}= 0.31\pm0.04$ of Ref.~\cite{Bass:2018uon}. 
However, systematic  uncertainties are associated with such extraction, in particular regarding what should be the highest $Q^2$ values acceptable for a Regge-type fit, and the assumption that the data are $Q^2$-independent.   
Thus, our projected results are expected to significantly improve, statistically and systematically, the intercepts values derived from photoproduction and low-$Q^2$ electroproduction.

\subsubsection{Lower energy run}

In addition to taking data at the highest CEBAF energy available, it is also beneficial to run with a lower energy beam. This allows to bridge the gap between the data we proposed to take and the existing world data. The gap is especially large for the neutron and it is clear from Fig.~\ref{fig:GDHrun_world} that in order to test accurately the GDH sum rule, closing this gap
is important. The Hall D low energy data will also greatly improve the world data quality and offer cross-check between two fully independent experiments. Finally, it provides an avenue to normalize our relative yield to obtain the absolute cross-section difference $\Delta \sigma$ in case of unforeseen issues in the determination of the normalization factor in Hall~D. 
Figures~\ref{fig:expect}, ~\ref{fig:expect_n}, and ~\ref{fig:expect_D}, show that 10 days of data taking shared between proton and deuteron at beam energy of e.g. 4\,GeV will provide sufficiently precise data to cross-check/normalize with the ELSA data.

\subsection{Sensitivity of the determination of the $\nu$-dependence of $\Delta \sigma$}\label{sec:ShapeSens}

It is important to recognize  that the choice of the fit form in Figs.~\ref{fig:expect}--\ref{fig:expect_D}
is only a working hypothesis based on the leading theory expectation. 
Interpreting the data with a Regge-based form is not a requirement.
The high precision of the data, the high-density binning and the large $\nu$-range will allow a clean extraction of the behavior of $\Delta \sigma (\nu)$ regardless of the actual theory driving its $\nu$-dependence.
This is illustrated by  the results in Table~\ref{table_ex} where various functional forms are used to fit the data shown in Fig.~\ref{fig:isospin} (generated assuming Regge behavior). As the $\chi^2$ values reveal, the other forms fail to fit the data satisfactorily for this isospin analysis.
(Again, the form $a\nu^b$ used in
Fig.~\ref{fig:isospin} is just a working hypothesis. The analysis can be carried out regardless of the actual $\nu$-dependence of $\Delta \sigma$).

\begin{table}[!h]
\caption{{\small Examples of functional forms used to fit the simulated data in Fig.~\ref{fig:isospin}, and resulting $\chi^2$/d.o.f. 
For the Regge case (second row) $\chi^2/\mathrm{d.o.f}\approx1$ by construction since this was the form assumed when simulating the data and bin-to-bin uncorrelated systematic uncertainties are expected to be negligible. 
The large differences between the various $\chi^2$/d.o.f. show that the proposed measurement precision and $\nu$ range are sufficient to determine accurately what type of functional form the data may follow.}}
\label{table_ex}
\vspace*{3mm}
\centering
\begin{tabular}{|c|c|c|} \hline
Fit form & $\chi^2$/d.o.f (isoscalar case) & $\chi^2$/d.o.f (isovector case) \\ 
\hline 
$a\nu^b$ & 1.0 & 1.2 \\
\hline 
$a+b\nu$ & 17.4 & 3.1 \\
\hline 
$a+b\nu+c\nu^2$ & 2.3 & 1.3 \\
\hline 	
$a+b\log\nu$ & 8.0 & 1.5 \\
\hline 	
$ae^{b\nu}+c$ & 3.8 & 5.0 \\
\hline 	
\end{tabular}
\end{table}

In contrast, if the $\nu$-range is reduced by half, e.g. to span only the region from 5 to 9.6\,GeV, then different functional forms may describe the data equally well, see Table~\ref{table_ex2}. 

\begin{table}[!h]
\caption{{\small Same as Table~\ref{table_ex} but with an experimental $\nu$-range reduced to half.}}
\vspace*{3mm}
\label{table_ex2}
\centering
\begin{tabular}{|c|c|c|} \hline
Fit form & $\chi^2$/d.o.f (isoscaler case) & $\chi^2$/d.o.f (isovector case) \\ 
\hline 
$a\nu^b$ & 1.1 & 1.0 \\
\hline 
$a+b\nu$ & 2.1 & 1.1 \\
\hline 
$a+b\nu+c\nu^2$ & 1.1 & 1.1 \\
\hline 	
$a+b\log\nu$ & 1.5 & 1.1 \\
\hline 	
$ae^{b\nu}+c$ & 1.3 & 1.2 \\
\hline 	
\end{tabular}
\end{table}
Table~\ref{table_ex} also demonstrates the added power of an isospin analysis compared to measuring only a single nucleon. Several fit forms can describe a single nucleon data adequately, but only the form chosen to generate the pseudo-data can fit both nucleon satisfactorily.

\section{Systematic uncertainties \label{sec:uncert_syst}}

\subsection{Uncertainties affecting the $\nu$ dependence of $\Delta \sigma$ \label{nu-dep errors}}

For studying the \textit{convergence} of the GDH integral, it is sufficient to obtain the high-$\nu$ behavior of the yield difference $\Delta y(\nu) =  N^+ - N^-$, and since the 
data at various $\nu$ are taken concurrently, an accurate absolute normalization of $\sigma_P-\sigma_A$ is irrelevant. 
Thus, the accuracy on this goal of the experiment depends only on the uncertainties affecting the $\nu$ dependence of $\Delta \sigma$.
It can be assessed with \emph{hdgeant4}, see Fig.~\ref{fig:hadron_effic}.
The tagger channel inefficiencies cancel in the flux normalization and thus do not contribute.

\subsection{Uncertainties affecting the absolute normalization of $\Delta \sigma$ \label{norm uncerts}}

For studying the \textit{validity} of the GDH sum rule, an absolute $\Delta \sigma$ is necessary.
Our primary method to obtain it will be by performing a standard cross-section analysis, except that A) the target dilution and unpolarized backgrounds need not to be corrected for as they do not affect the cross-section difference. (The very small polarized background contribution can be corrected for, see Section~\ref{sec:simulation}), and B), no knowledge of the target density is necessary due to a ratio cancellation between target polarimetry and absolute cross-section normalization. Hence, target density uncertainty does not contribute to the total uncertainty.
The systematics uncertainties associated with this method are:
\begin{itemize}
\item Beam polarization: $\delta P_e = 3\%$,  with 2\% due to precession and knowledge of beam energy, 1\% due synchrotron radiation depolarization and 1\% from Mott/Hall polarimeters.
\item Target polarization (without target density uncertainty contribution): $\delta P_t = 3\%$.
\item The photon flux uncertainty, $\delta \phi < 1\%$.
\item The combination of absolute detector, trigger and DAQ efficiencies
is assumed to be known to within 2-3\%.  
\end{itemize}
This yields an estimated total systematic uncertainty of 5.0\%.

The absolute $\Delta \sigma$ can alternatively be obtained by measuring the asymmetry $A=(N^+ - N^-)/(N^+ + N^-)$ and using the well-measured unpolarized cross-section $\sigma$ to provide $\Delta \sigma =2 \sigma A$. We can use the quicker relative asymmetry method as an on-line analysis method and later as a check of the primary method. We assume the following values for uncertainties:

\begin{itemize}
\item Electron beam polarization: $\delta P_e = 3\%$.
\item Target polarization (including target density uncertainty contribution)): $\delta P_t = 4\%$.
\item Target dilution: $\delta D = 3\%$.
\item Unpolarized cross-section $\sigma$ (from world data): $\delta \sigma = 1\%$.
\end{itemize}
This yields a total uncertainty of 5.9\% slightly larger than the absolute $\Delta \sigma$ analysis, but comparable.

\subsection{Uncertainties and target spin flip}

Possible false asymmetries related to the beam and target polarizations can be minimized by flipping the target spin and reversing the beam helicity assignment with the beam half-wave plate.
There is no single-spin longitudinal asymmetry for photoproduction reactions (in contrast to electroproduction). 
A non-uniform acceptance of the apparatus in the polar direction may induce a bias in the data if the $\sigma_P$ and $\sigma_A$ cross-sections have different polar angle dependence.  
Reversing the target spin once during the experiment to get two data sets of opposite raw asymmetry sign will ensure than the above asymmetry (and other possible ones) cancels when the two data sets are combined.
Reversing the target spin is relatively easy and fast (\sfrac{1}{2} day) and it will add robustness to the final result.

\section{Proposed schedule and beam time request}\label{sec:request}

Table~\ref{tab:beamtime} summarizes the proposed schedule and beam time request.
The incentive for starting the experiment with the deuteron is two-fold: 
1) it is faster to switch from deuteron to proton (12h) than the reverse (36h); and
2) a spin dance to confirm that the beam precession angle is known should be done as early as possible, and the deuteron asymmetry is expected to be larger than that of the proton. 

To obtain a comparable statistical precision for the proton and neutron requires the deuteron run time to be $\sim \sqrt2$ longer than that of the proton in order to account for subtraction of the proton.
For the 12 GeV run, operating 7 days on the proton target and 10 days on the deuteron will yield a relative statistical uncertainty on the 
coefficients $c_1$ and $c_2$ of Eq.~(\ref{eq:Regge_expect}) of about 5\%. This is comparable to the expected systematic uncertainty of about 5\% on the absolute cross-section, thereby making 7+10 days an optimal run time for the measuring the absolute $\Delta \sigma$.
(For the other main goal of the experiment --the intercept measurements-- the systematics uncertainty is negligible and their precision will be given by the 2-4\% statistical uncertainty.)
We also request 4 days (proton) + 5.7 days (deuteron) of beam time for the low beam energy run.

 To minimize the systematic uncertainties, the target spin will be flipped (12h) once for each target.  The target will be repolarized to its optimal value during the spin-flip process and the target NMR polarimetry recalibrated (additional 12h).
For the absolute cross-section determination, it is necessary to calibrate each of the three Pair Spectrometer configurations that would be necessary to span the full energy range of the tagger.  This would require three 4-hour ``TAC" runs.
Finally, to allow for the possibility of an asymmetry analysis---for which, in contrast to $\Delta \sigma$, the unpolarized background needs to be corrected for---another 0.5 day of empty target data taking at 12 GeV and a 0.3 day at 4 GeV are necessary.
In all, the above program requires 29.1 days of beam and 4 days without beam for target and beam configuration changes.

\begin{table*}[ht]
\caption{{\small Beam time requested and overhead, listed chronologically.
The beam current for production is 240 nA.}}
\label{tab:beamtime}
\begin{center}
\begin{tabular}{|c|c|c|} \hline
Time (day)          &Target                &Goal/Remarks   \\ \hline              
10    &Deuteron      & Main production at 12 GeV           \\   \hline          
0.3    &Deuteron      & Spin dance done during above task \\    \hline        
1    &Deuteron      & Target spin-flip/repol./NMR calib. \\
       &              &No beam, done at middle of production  \\  \hline     
0.5    & $^4$He      &  For background subtraction. \\
     &          &  Includes target change overhead \\\hline  
1   &Deteuron $\to$ proton switch     & No beam. NMR calib.           \\  \hline      
7      & Proton         & Main production at 12 GeV             \\  \hline          
1    &Proton      & Target spin-flip/repol./NMR calib. \\
       &              &No beam, done at middle of production  \\  \hline 
0.5      & Pair. Spec. converter    & Absolute flux calib.             \\  \hline
{\bf{12 GeV: 21.3 }}         & &     total time at 12 GeV     \\ \hline   \hline

5.7      &Deuteron           &Production  4 GeV      \\  \hline         
0.3    &Deuteron      & Spin dance done during above task \\    \hline       
0.3    & $^4$He      &  For background subtraction. \\
     &          &  Includes target change overhead \\\hline  
1   &Deuteron $\to$ proton switch.     & No beam. NMR calib.           \\  \hline  
4       &Proton            &Production at 4 GeV              \\ \hline 
0.5      & Pair. Spec. converter    & Absolute flux calib.             \\  \hline 
{\bf{4 GeV: 11.8 }}         & &    total time at 4 GeV      \\\hline 
{\bf{Total: 33.1 }}         & &     total experiment time    \\ \hline   
\end{tabular} 
\end{center}
\end{table*}

\section{Impact of the results}\label{sec:impact}

Studying the convergence properties of the GDH integral in the Regge ($\nu>3$\,GeV) domain is a first goal of the experiment that can be quickly and reliably reached. Then, once an absolute cross-section analysis is carried out, the accuracy at which the sum rule is tested for both the proton and the neutron can be revisited and significantly improved.
In addition to this, the proposed $\Delta\sigma(\nu)$
data at high $\nu$ will improve our knowledge on both the imaginary 
and real parts of the spin-dependent Compton amplitude $f_2$; 
it will provide new information on the poorly known intercept of the $a_1$ Regge trajectory; 
it will yield the first non-zero polarized deuteron asymmetry in the diffractive regime (assuming current predictions for the nucleon polarized rates in that regime), thereby providing for the first time a non-zero value for the isosinglet coefficient of $\Delta\sigma$;
it will reduce the uncertainty of the polarizability contribution to $1S$ hyperfine  splitting of hydrogen; and it will provide a photon-point benchmark to study the transition between the well-understood DIS dynamics of QCD to the lesser-known dynamics of diffractive scattering that will be explored with the EIC~\cite{Accardi:2012qut}.
These seven items are discussed separately in the following.

\subsection{Convergence of the GDH integral}

If it is found that the data obey the Regge theory, $\Delta\sigma \propto \nu^b$, in the measured $\nu$-range, one can extrapolate this behavior to larger $\nu$. The integral will converge if $b < 1$ (for both isoscalar and isovector components). 
With the expectation that $|b|\approx0.5$ and the capability of the experiment to determine it on a few percents level, 
the question of the convergence of the GDH integral will be settled.

As emphasized in Section~\ref{sec:uncert_stats}, the assumption of a Regge behavior is only a working hypothesis and other behaviors can be chosen if the data do not obey the Regge expectation.
Since no structure in the $\nu$-dependence of $\Delta\sigma(\nu)$  is expected above 3\,GeV, a definitive statement on the convergence is expected regardless of whether the data obey the Regge expectation or not. 
We emphasize that the finding that Regge theory fails in the spin sector would be very significant by itself.

\subsection{Verification of the GDH sum rule on the proton and neutron}

\subsubsection{Proton}

Assuming the validity of the Regge theory (or alternatively of the GDH sum rule)
we expect to measure between 3 to 12\,GeV a contribution to the proton GDH integral $I_p$ of about $-20\,\mu$b with negligible statistical uncertainty and a $1.0\,\mu$b systematic uncertainty, see Section~\ref{norm uncerts}. 
This would change the current assessment of $I_p$ from:\\
$
\big(226\pm 5$\,(stat) $\pm12$\,(syst) $\pm10$\,(large-$\nu$ projection)$\big)\,\mu$b
to \\
$\big(205\pm 5$\,(stat) $\pm12$\,(syst) $\pm1$\,(large-$\nu$ projection)$\big)\,\mu$b.\\ 
Thus, the total systematic uncertainty will decrease from 16 $\mu$b to 12 $\mu$b. The precision of the sum rule will be reduced from $16/205=8\,\%$ to $12/205=6\,\%$,  
a relative improvement of 25\% on the precision at which the GDH sum rule for the proton is currently tested.

\subsubsection{Neutron}
There is presently no assessement on the validity of the GDH sum rule on the neutron. Our data complementing those of MAMI and ELSA will offer the first test, with a precision comparable to that of the proton.

\subsection{Determination of the real and imaginary parts of the spin-dependent Compton amplitude $f_2(\nu)$ \label{impact with f2}}

\begin{figure}[!hbtp]
\begin{center}
\includegraphics[width=9.5cm]{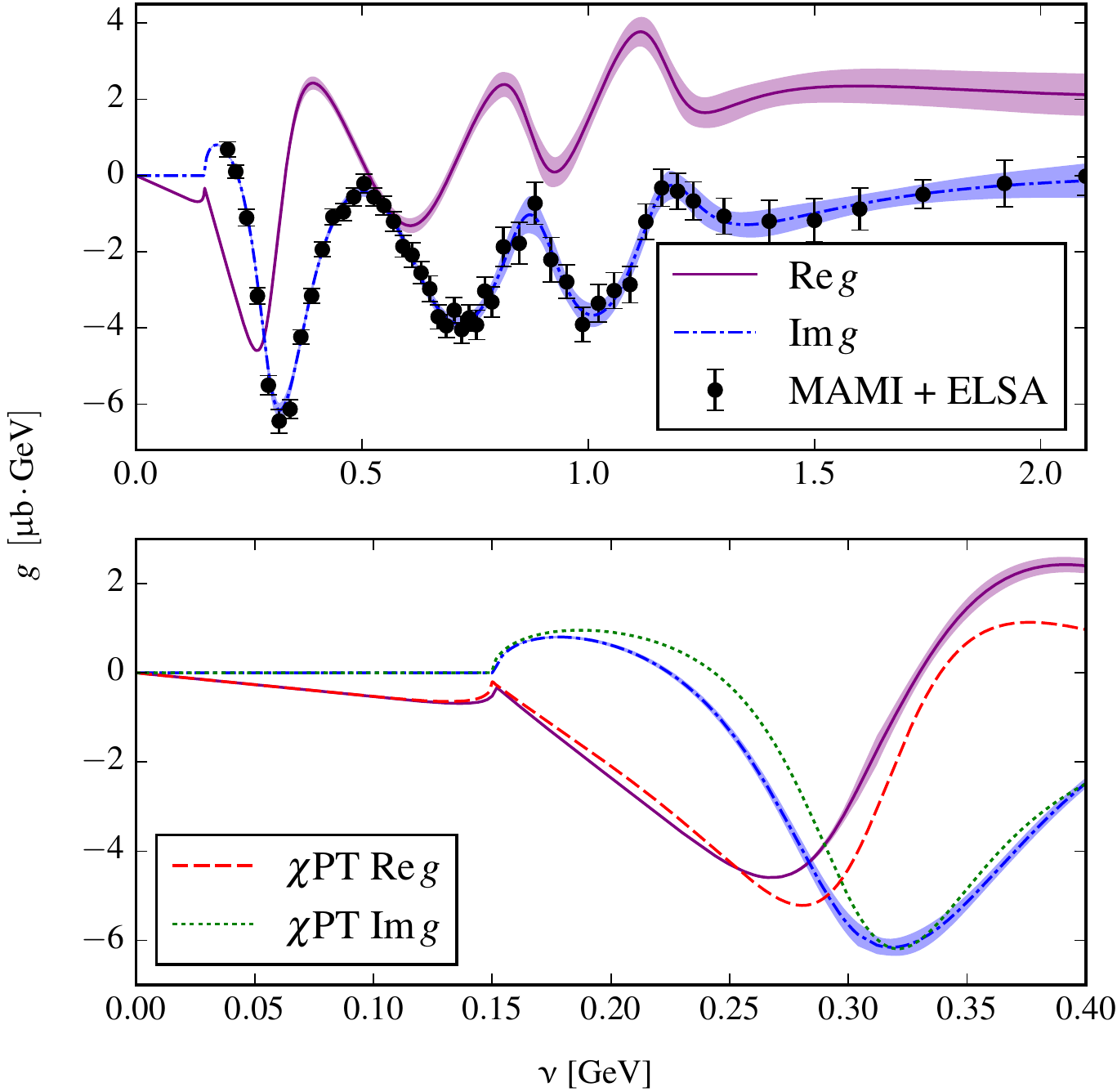}
\end{center}
\vspace*{-0.4cm}
\caption{\small{The proton spin-dependent Compton amplitude $f_2(\nu)$, denoted $g$ in the figure.  Top: real and imaginary parts, the latter fitted to GDH data, 
the former calculated via dispersion relations. Bottom: comparison to NNLO $\chi$EFT calculation at low $\nu$ indicating that the measured (blue band) and calculated (dotted green line) imaginary parts differ appreciably at energies around 0.25 GeV while the real
parts (obtained by integrating the imaginary parts over
the energy domain with $\nu$ as the integration parameter) 
agree perfectly at low $\nu$.
This reflects the peculiar feature of the theory that low-energy
quantities are well described, even though they are obtained 
as loop or dispersive integrals which include higher-energy
domains where the theory is inapplicable.
Figure from~\cite{Gryniuk:2016gnm}.}}
\label{fig:g}
\end{figure}

The spin-dependent Compton amplitude $f_2(\nu)$, also denoted 
by $g(\nu)$ in literature, is a complex quantity whose imaginary part 
is determined by $\Delta\sigma$, see Eq.~(\ref{eq:optheo}), and will 
thus be measured directly by the experiment.  Fig.~\ref{fig:g} 
(top) shows the world data on $\Im m(f_2)$ for the proton, extracted 
from $\Delta\sigma$ measured at MAMI and ELSA.  

It is a lovely feature of the proposed experiment that it allows us to access Compton physics and helps us to constrain other pertinent 
unpolarized and polarized observables without resorting to a dedicated 
Compton setup.  Specifically, once $\Im m(f_2)$ is obtained from $\Delta\sigma$, the real part of the spin-dependent amplitude, 
$\Re e(f_2)$, can be determined from $\Im m(f_2)$ by using 
Eq.~(\ref{disprel}) \cite{Hagelstein16}.  The reliability of this extraction is shown by the violet error band in Fig.~\ref{fig:g}, and strongly depends on the quality of $\Im m(f_2)$ (blue error band).  
It is clear that both error bands increase as $\nu$ reaches the upper portion of the previously covered energy region, and will continue to do 
so at higher $\nu$ unless high-quality data will be made available.  
Our
data will extend 
the $\nu$-coverage and permit this symbiosis of $\Re e(f_2)$ 
and $\Im m(f_2)$ to six times its present reach.

\clearpage

\begin{figure}[hbtp]
\begin{center}
\includegraphics[width=10cm]{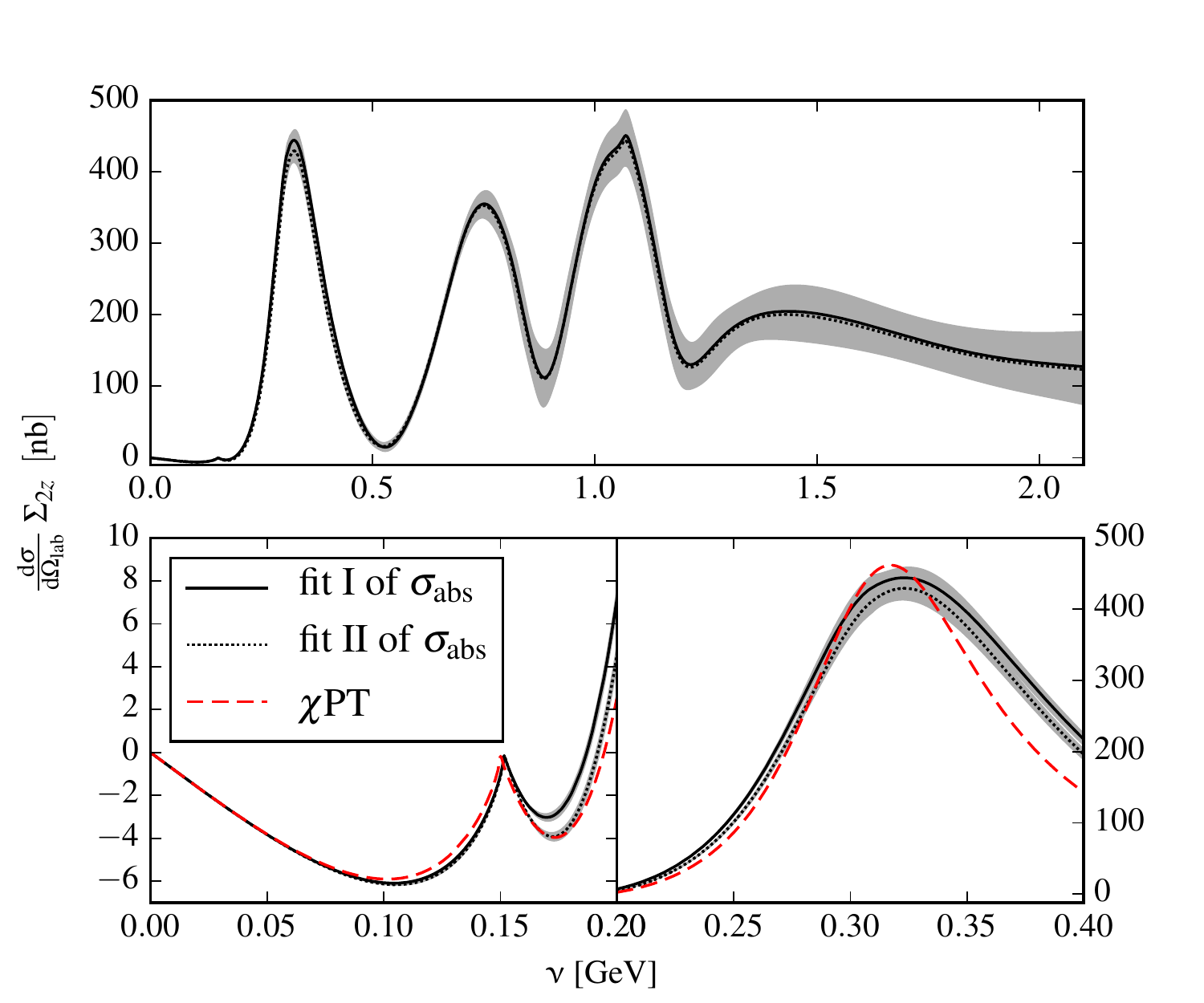}
\end{center}
\vspace*{-4mm}
\caption{\small{Unpolarized differential cross-section multiplied with
the $\Sigma_{2z}$ asymmetry for the forward Compton scattering
off the proton, showing (top) two distinctive fits of the unpolarized 
photoabsorption cross-section and its uncertainties, and (bottom)
the $\chi$EFT calculation.  Figure from \cite{Gryniuk:2016gnm}.}}
\label{fig:2refg}
\end{figure}

If both $\Re e(f_2)$ and $\Im m(f_2)$ are known precisely enough 
(and given $f_1$, which is well measured), the two complex amplitudes 
can be used to determine $\mathrm{d}\sigma/\mathrm{d}\Omega$ 
and the beam-target asymmetry $\Sigma_{2z}$ in the forward limit, i.e.,
$$
\left. {\mathrm{d}\sigma\over \mathrm{d}\Omega} \right\vert_{\theta=0}
  = \bigl\vert f_1 \bigr\vert^2 + \bigl\vert f_2 \bigr\vert^2 \>, \qquad
\left. \Sigma_{2z} \right\vert_{\theta=0}
  = - {2\Re e (f_1 f_2^*) \over |f_1|^2 + |f_2|^2} \>,
$$
where $\theta$ is the Compton scattering angle and $\overrightarrow{z}$ 
is along the initial photon direction.  Of these,  
$\Sigma_{2z} |_{\theta=0}$ is most interesting since the asymmetry 
for circularly polarized photons and nucleons polarized along the $z$ axis,
$$
\Sigma_{2z} 
  = {\mathrm{d}\sigma_P - \mathrm{d}\sigma_A \over
     \mathrm{d}\sigma_P + \mathrm{d}\sigma_A} \>,
$$
provides information on all four spin polarizabilities appearing 
in Compton scattering.  In particular $\Sigma_{2z}$ and its behavior
near $\theta=0$ are very sensitive to chiral loops \cite{Lensky15}.  
The product of the unpolarized cross-section and $\Sigma_{2z}$ 
for $\theta=0$ is shown in Fig.~\ref{fig:2refg} (top) together with 
its uncertainty, which increases rapidly for $\nu \gtrsim 2\,\mathrm{GeV}$.
The precise measurement of $\Delta\sigma(\nu)$ in the $\nu$ range covered 
in Hall D will significantly reduce the uncertainty on $\Sigma_{2z}$.

The analysis~\cite{Gryniuk:2016gnm} was performed for the proton only. In addition to improving 
it with our higher-$\nu$ high-precision proton data, our neutron and deuteron data will 
motivate the same type of analyses for these objects. 
$\chi$EFT is an important effective approach to QCD that should describe it at low 
energies and momenta.  However, the dedicated JLab low $Q^2$ experimental program 
to test $\chi$EFT with spin observables is showing that their description is a challenge 
to $\chi$EFT \cite{Deur:2018roz}.  Thus, providing further tests of $\chi$EFT with new 
spin observables or/and in a different regime is critical and can be achieved with 
the present proposal.

\subsection{The intercept of the $a_1$ Regge trajectory \label{impact on a_1 intercept.}}

In Regge theory, the high-energy behavior of the isovector (non-singlet)
cross-section difference 
is driven by the $a_1(1260)$ 
Regge trajectory such that
\begin{equation}
    \Delta\sigma^{(p-n)} 
  \sim s^{\alpha_{a_1}-1} \> ,
\end{equation}
where typically $\alpha_{a_1} \approx 0.4$, 
as obtained from fits to DIS  data.
A very recent such fit~\cite{mvdh19} resulted in $\alpha_{a_1} \approx +0.45$ 
while 
the Regge expectations is $\alpha_{a_1} \approx -0.34$ 
(see Eq.~(\ref{eq:Regge_expect_alphas}) below).
Another recent fit, combining both electroproduction and photoproduction data~\cite{Bass:2018uon}, yields $\alpha_{a_1} = +0.31\pm0.04$, i.~e.~also finds that the sign of the $a_1(1260)$ intercept is opposite to the theoretical prediction.  The situation is summarized
in the Table below.
\begin{center}
\begin{tabular}{ll}
\hline

\hline

\hline
 & $\phantom{-}\alpha_{a_1}$  \\
\hline
DIS fit (approx. values) & $\phantom{-}0.45$  \\
Photo/electro-production fit & $\phantom{-}0.31\pm 0.04$ \\ Regge expectation & $-0.34$ \\
\hline

\hline

\hline
\end{tabular}
\end{center}
The problem at the root of the discrepancy on $\alpha_{a_1}$ is partly that $a_1(1260)$ is the only $I^G(J^{PC}) = 1^-(1^{++})$ meson to form a ``trajectory'', while the second candidate, the $a_1(1640)$, has been omitted from the PDG Summary Tables as it still needs confirmation.
A precise measurement of $\Delta\sigma$ at high $\nu$ for both proton and neutron targets would help to remove this uncertainty.
This is an important question to resolve as the intercept is predicted to be given by
\begin{equation}
\alpha_{a_1} = 1 - \alpha' m_{a_1}^2 \>,
\label{eq:Regge_expect_alphas}
\end{equation}
where
$\alpha' = 1/(2\pi\sigma) \approx 0.88 \, \mathrm{GeV}^{-2}$
and $\sigma$ is the string tension, which is known to be approximately
$0.18 \,\mathrm{GeV}^2$.  
If $\alpha_{a_1}$ were indeed $\approx 0.45$ as suggested by the present DIS data and the (relatively low-$\nu$) photoproduction data,
this would imply 
$\alpha' \approx 0.44 \, \mathrm{GeV}^{-2}$
and a string tension more than twice as high as the value commonly accepted and obtained from hadron spectroscopy.

As shown in Section~\ref{sec:uncert_stats}, if $\Delta \sigma$ obeys the presumed Regge behavior,  the experiment would determine $\alpha_{a_1}$ at a 
level of 2\%, an improvement in precision of a factor of 25 
compared to the present 54\% uncertainty obtained from the best fit 
to the world data.

\subsection{Deuteron asymmetry \label{Deuteron asymmetry}}

Since only null asymmetries have been measured by COMPASS, CLAS and SLAC for the deuteron in the low $Q^2$, high-$\nu$, regime relevant to Regge theory, the deuteron coefficient $2c_2$ (see Eq.~(\ref{eq:Regge_expect})) that factors the $s$-dependence of  $\Delta \sigma^{p+n}$ is assumed to be zero in analyses~\cite{Bass:2000zv, Bass:2018uon}. 
From the Regge expectation, a non-zero deuteron asymmetry, i.~e. $\Delta \sigma^{p+n} \neq 0$, should be unambiguously measured by this experiment, see Fig.~\ref{fig:expect_D}, yielding a clear non-zero $2c_2= 450 \pm 34$.

\subsection{Polarizability correction to hyperfine splitting in 
hydrogen \label{impact HFS}}
A valuable impact of the measurement concerns the effect of proton structure on the hyperfine splitting in hydrogen. The importance of this topic has been emphasized by the ``proton radius puzzle''~\cite{Pohl:2013yb}.  The hyperfine splitting is given by
\begin{equation}
E_\mathrm{HFS}(nS) 
  = \left[ 1 + \Delta_\mathrm{QED} + \Delta_\mathrm{weak} 
             + \Delta_\mathrm{structure} \right] E_\mathrm{Fermi}(nS) \>,
\end{equation}
where the proton-structure correction can be separated into three terms: 
the Zemach radius, the recoil contribution, and the polarizability contribution:
\begin{equation}
\Delta_\mathrm{structure}
  = \Delta_Z + \Delta_\mathrm{recoil} + \Delta_\mathrm{pol} \>.
\end{equation}
The current relative uncertainties of the three terms are 140~ppm, 0.8~ppm and 86~ppm, respectively, which need to be put into the perspective of the forthcoming PSI measurement of $E_\mathrm{HFS}$ whose precision is expected to be as low as 1~ppm.  
Our proposed measurement can contribute to the uncertainty reduction 
of $\Delta_\mathrm{pol}$.  It can be written as
\begin{equation}
\Delta_\mathrm{pol}
  = {\alpha_{em} m \over 2 \pi ( 1 + \kappa) M} [ \delta_1 + \delta_2 ] \>,
\end{equation}
where $m$ is the lepton mass (muon in the case of muonic
hydrogen where the effect is easiest to measure).  
Here $\delta_1$ involves an integral of the spin structure 
function $g_1(x,Q^2)$ over both $x$ and $Q^2$, 
\begin{equation}
\delta_1
  = 2 \int_0^\infty {\mathrm{d}Q\over Q} \biggl( 
    \biggl\{ \cdots \biggr\} 
    + {8M^2\over Q^2} \int_0^{x_0} \mathrm{d}x \, g_1(x,Q^2)
    \biggl\{ \cdots \biggr\} \biggr) \>,
    \label{delta1g1}
\end{equation}
while $\delta_2$ involves a similar integration 
of $g_2(x,Q^2)$ (see Eq.~(6.43b) of \cite{Hagelstein16}
for full expressions).  The GDH integrand at general
values of $\nu$ and $Q^2$, expressed in terms of 
the polarized nucleon structure functions $g_1(\nu,Q^2)$ 
and $g_2(\nu,Q^2)$, is
\begin{equation}
\Delta\sigma
  = - {8\pi\alpha_{em}^2\over M(\nu-Q^2/2M)}
    \left( g_1(\nu,Q^2) - {Q^2\over \nu^2} \, g_2(\nu,Q^2) \right) \>.
\end{equation}
At low $Q^2$ (and for real photons), $g_2$ is irrelevant,
hence a precise measurement of $\Delta\sigma$, such as
it will be provided by our experiment, directly constrains $\delta_1$ via $g_1$.  To compute $\delta_1$, one indeed needs 
the $Q^2$-dependence of $g_1$, i.~e.~use input from electron-scattering, but since the integrand, as seen in Eq.~(\ref{delta1g1}),
is weighted by $1/Q^3$, knowing the value at $Q^2=0$ from our real-photon measurement would be extremely beneficial in stabilizing 
the integration.
Such a stabilization is essential, as the 86~ppm uncertainty 
mentioned above needs to be reduced to $\approx 1$~ppm. This
implies that our knowledge of $g_1$ needs to be improved by two orders of magnitude.

\subsection{Transition between polarized DIS and diffractive regimes \label{impact EIC}}

As already quoted from Ref.~\cite{Drechsel:2007sq},
\emph{``above the resonance region [...] the real photon is essentially absorbed by 
coherent processes, which require interactions among the constituents such as gluon exchange
between two quarks. This behavior differs from DIS, which refers to incoherent scattering off the constituents.''}
That is, there is a transition between the DIS regime and the very low-$x$ or real photon 
regimes of diffractive scattering.
Studying this transition has been an important part of the ZEUS and H1 programs at HERA, and it remains a
very active field of research~\cite{TOTEM}. However, it is currently limited to unpolarized scattering.
The polarized case and its connection to photoproduction is discussed in Ref.~\cite{Bass:2000zv}
and will be explored with the EIC~\cite{Accardi:2012qut}. 

\begin{figure}[h!]
\center 
\includegraphics[scale=0.27]{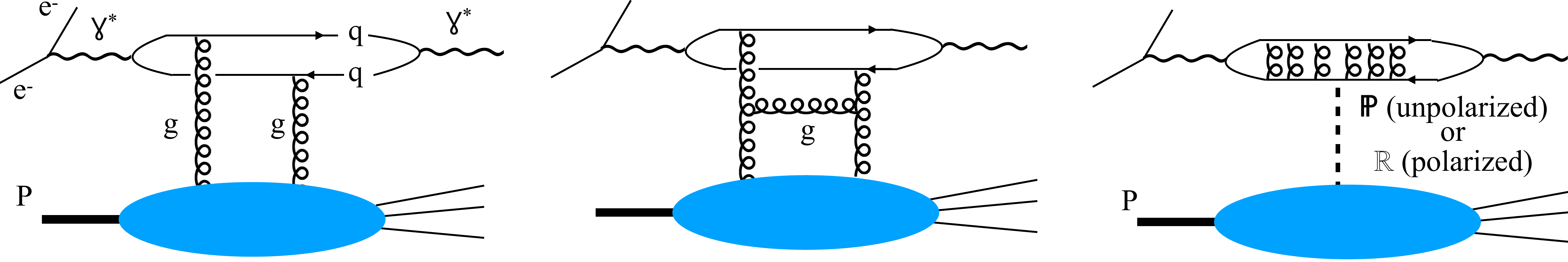}
\caption{\small{\label{fig:diff} Diquark picture of low-$x$ electron-proton scattering, 
from the higher $Q^2$ hard regime (left)
to the low $Q^2$ soft regime with Pomeron or Reggeon exchange (right).}}
\end{figure}

\begin{wrapfigure}{r}{0.35\textwidth} 
\begin{center}
\includegraphics[scale=0.25]{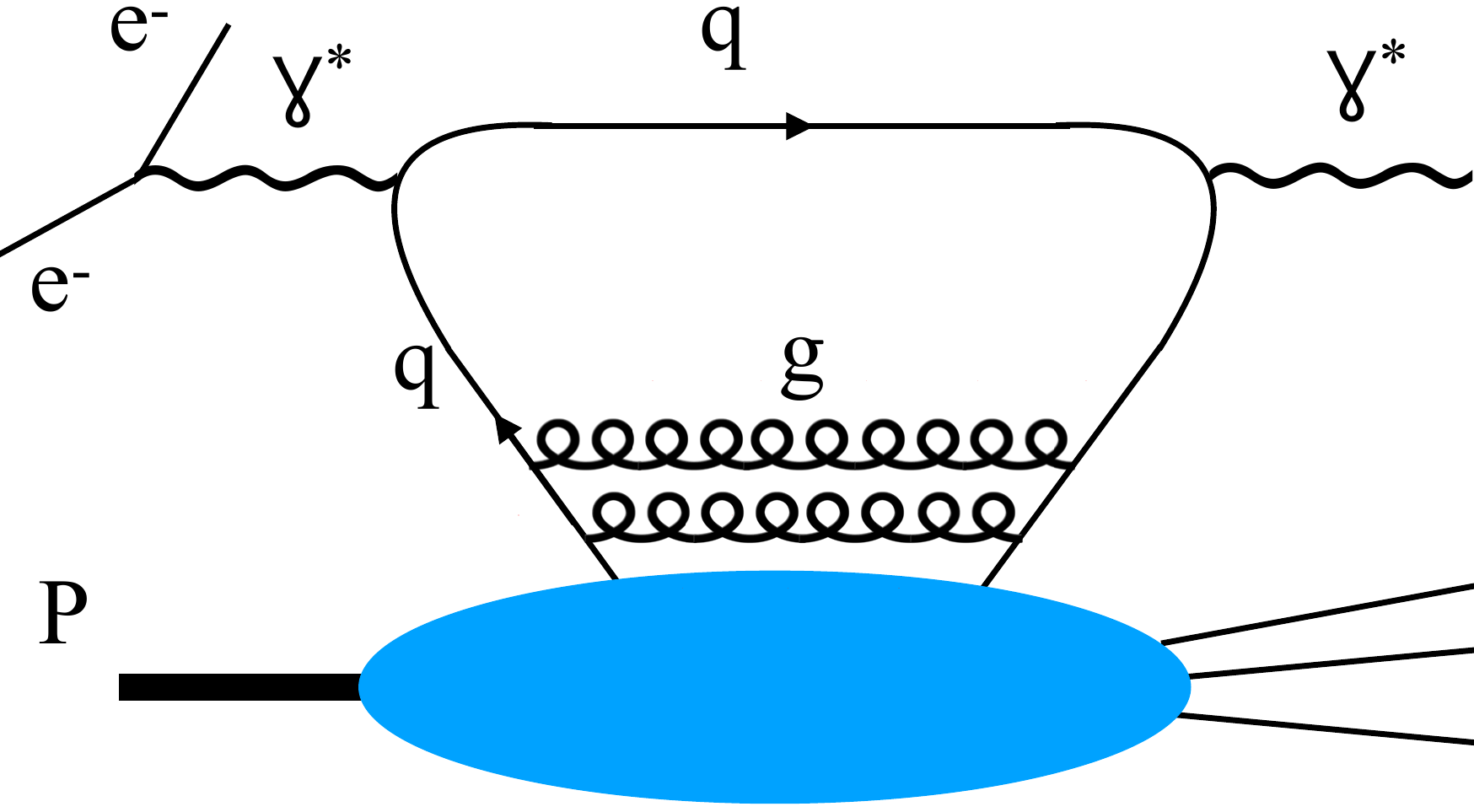}
\end{center}
\caption{\small Another possible process contributing to electron-proton scattering.}
\label{fig:diff2}
\end{wrapfigure}
The usual theoretical description of diffractive scattering 
is the diquark picture: the hard virtual photon emitted by the scattered 
lepton hadronizes into a $q \bar{q}$ pair of coherent length $1/(xM$). 
At high enough $Q^2$, each quark exchanges a gluon with the 
proton, see Fig.~\ref{fig:diff}, left panel. As $Q^2$ decreases,
gluon rungs on the gluon ladder appear (Fig.~\ref{fig:diff}, central panel), as well as gluons exchanged between the $q$ and $\bar{q}$. At low $Q^2$, the interaction between the coherent $q \bar{q}$ pair
and the proton is summed into pomeron ($\mathbb{P}$) and reggeon ($\mathbb{R}$) exchanges (Fig.~\ref{fig:diff}, right panel). Other processes contributing to $\mathbb{P}$ and $\mathbb{R}$ exchanges
exist, such as the one shown in Fig.~\ref{fig:diff2}. This description 
connects to the usual DIS parton model, e.~g.~with 
the gluons in the left panel of Fig.~\ref{fig:diff}
representing the gluon PDF. 

The pomeron has the vacuum quantum numbers (isoscalar charge singlet). Being spin 0 allows $\mathbb{P}$ to couple to the proton components irrespective of their helicity.  $\mathbb{P}$ thus controls 
unpolarized diffractive scattering.  Doubly polarized $\overrightarrow{e'}\overrightarrow{P}$ scattering filters out $\mathbb{P}$ exchanges to reveal the non-singlet $\mathbb{R}$ exchange. 
This filter will be used for the first time at the EIC.  This proposed measurement of $\Delta \sigma$, expected to be also controlled by Regge theory, will provide a  $Q^2=0$ baseline to this study of the transition from the hard dipole partonic picture to the soft $\mathbb{R}$ exchange picture.

\section{Summary}

We propose the first measurement of the high-energy behavior of the integrand $\Delta \sigma/\nu$ of the
GDH sum rule, a fundamental relation of quantum field theory whose validity depends on the internal
dynamical properties of the particle to which the sum rule is applied.
The measurement would be performed in Hall D, the only place suited for carrying out a high energy GDH measurement, using a FROST target and a longitudinally polarized electron beam on an aluminum radiator. 
The high-$\nu$ domain is where the sum rule may fail. In fact, the unpolarized
equivalent of the GDH integral does not converge, both for proton and neutron. This could be observed only
from high-$\nu$ data, $\nu>3$\,GeV, which is greater than the upper reach (2.9 GeV for the proton, 1.8 GeV for the neutron) of existing measurements of the GDH integrand. 
The proposed measurement, up to  $\nu = 12$\,GeV, would allow us to study the convergence property of the GDH integral. 
This can be achieved by a quick and robust analysis since unpolarized backgrounds cancel in $\Delta \sigma$ and no absolute normalization is needed. Then, once the absolute normalization is determined, the Hall D data added to the world data at lower energy will make a relative improvement of 25\% on the accuracy  at which the sum rule is tested on the proton, and provide for the first time a test of similar accuracy for the neutron.

In order to fill a large gap of missing neutron and deuteron data, it is necessary to perform a shorter measurement at lower energy, with the beam energy between 4\,GeV and 6\,GeV.  This is required to extract the neutron integral without the use of a model or extreme interpolation.  This will also allow us to precisely determine the minimum energy at which the Regge phenomenology is valid, as well as to constrain systematic uncertainties related to relative polarization of the photon beam and tagger geometry.  It will  provide an overlap between the Hall D data and the existing world data and allow $\Delta\sigma$ to be significantly improved in the higher resonance region for the proton.

In addition to studying the convergence and sum rule validity and independent of that study conclusion, the data will constrain our knowledge
of diffractive QCD, whose phenomenology is unverified in the spin sector. As pointed out in~\cite{Pantforder:1998nb},
not even a model prediction is available for the magnitude of the $J=1$ pole effect, due to our absence of knowledge 
of polarized diffractive QCD. In fact, results from fits of photoproduction data and of DIS data independently disagree
with the Regge theory expectation for the sign of the Regge trajectory intercept driving the isovector part of $\Delta \sigma$. 
The experiment will clarify this problem.

Given the Regge theory expectation, the experiment should measure the first non-zero asymmetry signal for the  
deuteron in the diffractive regime, thereby providing for the first time a non-null determination of 
the coefficient that factors the $s$-dependence of  $\Delta \sigma^{p+n}$ of the deuteron.

Analyzing $\Delta \sigma(\nu)$ using dispersion relation techniques will provide $f_2(\nu)$, 
the spin-dependent forward Compton amplitude. This will further clarify the convergence property
of the GDH sum rule, which depends on both the real and imaginary parts of $f_2$, and will test 
$\chi$EFT. The latter is especially important since tests of $\chi$EFT with polarized observables 
by the JLab low-$Q^2$ spin sum rule experimental
program revealed that currently, $\chi$EFT has difficulties to consistently describes 
spin observables~\cite{Deur:2018roz}. 

Furthermore, the experiment will provide a $Q^2=0$ baseline for the EIC data. This will
be helpful in particular for  the study of the transition between the DIS regime characterized 
by partonic degrees of freedom to the
diffractive regime characterized by effective degrees of freedom such as the pomeron and the reggeon. 

Finally, the data will constrain the polarizability contribution to the 
hydrogen hyperfine splitting.

A first goal of the experiment is to map with high precision the energy dependence of $\Delta \sigma$ on the proton and neutron.  
This will determine whether  $\Delta \sigma$ follows the expected Regge behavior and if so, the values of the isovector and isoscalar Regge trajectory intercepts will determine if the integral converges. 
Only point-to-point uncorrelated errors contribute to the Regge intercept uncertainties, which guaranties a fast and robust analysis.
Other goals for the proposal require absolute normalization.
The necessary information (e.g. polarization) will be gathered concurrently.
However, the convergence test does not require the absolute normalization and thus will have much reduced uncertainties compared to the absolute measurement.

With 27 days of measurement (10 days on deuteron at 12 GeV and 5.7 days at 4 GeV, and one week on proton at 12 GeV and 4 days at 4 GeV) plus 6 days for systematic studies and target changes, and assuming that Regge behavior is observed, the data will provide the Regge trajectory intercepts at the 2--4\% level, compared to the 50\% uncertainties at which they are presently known.

Once a polarized target is available in Hall D, a rich experimental program will open.
For example, several possible experiments have been discussed in an earlier LOI~\cite{Keller_LOI}. A GDH experiment would initiate such a program with a comparatively simple set-up and robust observables.

\end{document}